\documentclass{article}

\usepackage[margin=1in]{geometry}
\usepackage[T1]{fontenc}
\usepackage[latin9]{inputenc}
\usepackage{amsmath}
\usepackage{amssymb}
\usepackage{amsthm}
\usepackage{xcolor}

\numberwithin{equation}{section}

\newtheorem{theorem}{Theorem}[section]
\newtheorem{lemma}[theorem]{Lemma}
\newtheorem{fact}[theorem]{Fact}
\newtheorem{corollary}[theorem]{Corollary}
\newtheorem{proposition}[theorem]{Proposition}
\newtheorem{definition}[theorem]{Definition}

\newtheorem{remark}[theorem]{Remark}

 \global\long\def\sbr#1{\left[ #1\right] }
 \global\long\def\cbr#1{\left\{  #1\right\}  }
 \global\long\def\rbr#1{\left(#1\right)}
 
 \global\long\def\E{\mathbb{E}}
 \global\long\def\P{\mathbb{P}}
 \global\long\def\R{\mathbb{R}}
 \global\long\def\Z{\mathbb{Z}}
 \global\long\def\N{\mathbb{N}}

 \global\long\def\dd#1{\textnormal{d}#1}

 \global\long\def\ra{\rightarrow}
 \global\long\def\TTV#1#2#3{\text{TV}^{#3}\!\rbr{#1,#2}}

 \global\long\def\ns{\infty}

\begin{document}

\title{BDG inequalities and their applications for model-free continuous price paths with instant enforcement}

\author{R. M. {\L}ochowski}

\maketitle

\begin{abstract}
Shafer and Vovk introduce in their book \cite{ShaferVovk:2018} the notion of \emph{instant enforcement} and \emph{instantly blockable} properties. However, they do not associate these notions with any outer measure, unlike what Vovk did in the case of sets of ''typical'' price paths. In this paper we introduce an outer measure on the space $[0, +\ns) \times \Omega$ which assigns zero value exactly to those sets (properties) of pairs of time $t$ and an elementary event $\omega$ which are instantly blockable. Next, for a slightly modified measure, we prove It\^o's isometry and BDG inequalities, and then use them to define an It\^o-type integral. Additionally, we prove few properties for the quadratic variation of model-free, continuous martingales, which hold with instant enforcement.
\end{abstract}

\section{Introduction}\label{intr}

Since the last subprime mortgage financial crisis there is a growing interest in the robust financial models, usually models with minimal, widely accepted non-arbitrage assumptions. Such assumptions together with game-theoretic considerations allow to establish properties which characterize trajectories of prices of financial assets which exclude possibility of arbitrage. In a series of papers, among others in \cite{Vovk_volatility:2008}, \cite{Vovk_randomness:2009}, \cite{Vovk_probability:2012},  \cite{Vovk_cadlag:2011}, \cite{Vovk_cadlag:2015}, Vovk introduced and considered outer measures on the spaces of continuous or more general, c\`adl\`ag trajectories, which assign zero value to the sets of trajectories of prices of financial assets, which allow for arbitrage. ''Typical'' (not leading to arbitrage) trajectories possess quadratic variation and model-free, It\^o-type integration with respect to such trajectories may be established (\cite{PerkowskiProemel_integral:2015}, \cite{Vovk_Schafer_semimart:2016}, \cite{LochPerkPro:2018}). 

The investigations in game-theoretic approach to model-free, financial models of continuous price paths culminated in Glenn Shafer and Vladimir Vovk publishing their book \cite{ShaferVovk:2018}. In their book Shafer and Vovk introduce a new notion -- the notion of \emph{instant enforcement}. But they do not characterize it using any outer measure, unlike what Vovk did in the case of sets of ''typical'' price paths. Informally, property $E$ is instantly enforceable if there exists a trading strategy making a trader using this strategy infinitely rich as soon as the property $E$ ceases to hold. In this paper we introduce an outer measure on the space $[0, +\ns) \times \Omega$, which assigns zero value exactly to those sets (properties) of pairs of time $t$ and an elementary event $\omega$, complements of which are instantly enforceable. We also introduce a slight modification of this measure (an open question is whether the introduced modification differs from the original measure) which allows us to establish It\^o's isometry and BDG inequalities for this modification. Such results were not present in Vovk or Shafer's works. A ''weak'' BDG inequality in a model-free setting, but quite different from ours (and only for $p=2$) was established in \cite{BartlKupperNeufeldSDE:2018}. A main novelty in our approach is that instead of working with (outer) expectation $\bar{\E}$ defined for variables $X:\Omega \ra [-\ns, +\ns]$, as for example in \cite[Sect. 13.3]{ShaferVovk:2018} or \cite{BartlKupperNeufeldSDE:2018}, we introduce a functional $\overline{\E}$ which is defined on (generalized) processes $X:[0, +\ns) \times \Omega \ra [-\ns, +\ns]$.
 Using the obtained BDG inequalities, we define an It\^o-type integral, which allows to integrate more general (not necessarily continuous) integrands than those considered in \cite{ShaferVovk:2018}. Finally, we present a sequence of processes, which do not depend on any partitions, which tends locally uniformly with instant enforcement to the quadratic variations of martingales. This is also a new result, not published elsewhere in the literature.

\subsection{Definitions and notation}
Now we outline a general setting in which we will work and which
follows closely \cite[Chapt. 14]{ShaferVovk:2018}. For simplicity, we consider
only finite families of basic martingales. We will work with a martingale
space which is a quintuple

\[
\rbr{\Omega,{\cal F},\mathbb{F=}\rbr{{\cal F}_{t}}_{t\ge0},J=\cbr{1,2,\ldots,d},\cbr{S^{j},j\in J}}
\]
of the following objects: $\Omega$ is a space of possible outcomes
of reality, whose elements are called \emph{elementary events}, ${\cal F}$ is a $\sigma$-field of the subsets of $\Omega$
which we call \emph{events}, $\mathbb{F}=\rbr{{\cal F}_{t}}_{t\ge0}$
is a filtration (writing $t \ge 0$ we mean that $t \in [0,+\ns)$) such that for $t \ge 0$, ${\cal F}_{t} \subseteq {\cal F}$, and $\cbr{S^{j},j\in J}=\cbr{S^{1},S^{2},\ldots S^{d}}$
is a family of mappings $S^j :[0, +\ns) \times \Omega \ra \R$, $j \in J$, called \emph{basic continuous martingales}, such that for any
$t \ge 0$ and $j\in J$, $S_{t}^{j}$ is a $\rbr{{\cal F}_{t},{\cal B}(\R)}$-measurable
\emph{real variable} $S_{t}^{j}:\Omega\ra\R$ (${\cal B}(\R)$ denotes the $\sigma$-field of Borel subsets of the set of real numbers $\R$) and such that for each $\omega\in\Omega$
the trajectory $[0,+\ns)\ni t\mapsto S_{t}^{j}(\omega)$ is continuous.

Throughout the paper the filtration $\mathbb{F}$ is fixed, moreover, we assume that ${\cal F}_{0}$ is trivial, ${\cal F}_{0}=\cbr{\emptyset,\Omega}$,
thus all $\rbr{{\cal F}_{0},{\cal B}(\R)}$-measurable variables $S_{0}^{j}$,
$j\in J$, are deterministic.
In the paper we need to work with some stopping times and therefore, to assure that the quantities we define are indeed stopping times with respect to $\mathbb{F}$, we make the following assumption.
\medskip
\newline
Assumption {\bf A}: \emph{for any $t \ge 0$ and any instantly blockable set $B\subseteq[0,+\ns)\times\Omega$ (blockable sets are defined is Sect. \ref{sect_defi}) the projection of $B \cap ([0,t] \times \Omega)$ onto $\Omega$ belongs to ${\cal F}_{t}$.}
\medskip
\newline
Assumption {\bf A} seems to be reasonable since it roughly means that at the moment $t \ge 0$ we are able to say if there was any trading strategy making us infinitely rich (after investing a small positive amount at the moment $0$) until the moment $t$. This assumption is similar to the frequently made assumption in a classical probabilistic setting that a filtration is complete. Alternatively, we may just assume that the set of basic martingales and the filtration $\mathbb{F}$ are such that all the times which we need to be stopping times are indeed stopping times (with resp. to $\mathbb{F}$).

\begin{remark}
{
A common way to assure that some debut or hitting times similar to those used in this paper are indeed stopping times, is to use the universal completion of $\sigma$-algebras (see for example \cite[p. 273]{Vovk_cadlag:2015}, \cite[p. 4081]{LochPerkPro:2018}). Since such an operation is complicated and has not obvious financial interpretation, we prefer to use assumption {\bf A}.}
\end{remark}

A \emph{real process} $X:[0,+\ns)\times\Omega\ra\R$ is a collection
of real variables $X_{t}:\Omega\ra\R$, $t \ge 0$, such that
$X_{t}$ is $\rbr{{\cal F}_{t},{\cal B}(\R)}$-measurable, thus all
processes which we consider are adapted to $\mathbb{F}$. 

A \emph{process} $Y:[0,+\ns)\times\Omega\ra\R\cup\cbr{-\ns,+\ns}=[-\ns,+\ns]$,
is a collection of \emph{extended variables} $Y_{t}:\Omega\ra[-\ns,+\ns]$,
$t\in[0,+\ns)$, such that $Y_{t}$ is $\rbr{{\cal F}_{t},{\cal B}([-\ns,+\ns])}$-measurable
(any set in ${\cal B}([-\ns,+\ns])$ is of the form $A$, $A\cup\cbr{-\ns}$,
$A\cup\cbr{+\ns}$ or $A\cup\cbr{-\ns,+\ns}$, where $A\in{\cal B}(\R)$). 

\emph{Any} mapping $Y:[0,+\ns)\times\Omega\ra[-\ns,+\ns]$ is called a \emph{generalized process}. Note that $Y_t$ does not need to be $\rbr{{\cal F}_{t},{\cal B}([-\ns,+\ns])}$-measurable and thus a generalized process may not be a process as defined in the previous paragraph.

For any generalized process $Y$ we define its supremum process $Y^*$, which
is a generalized process defined as
\[
Y_{t}^{*}(\omega):=\sup_{0\le s\le t}\left|Y_{t}(\omega)\right|,
\]
where we denote $Y_{t}(\omega):=Y(t,\omega)$. 

A generalized process $Y$ is \emph{globally bounded} if 
\[\sup_{(t,\omega) \in [0, +\ns) \times \Omega} |Y_t(\omega)| = \sup_{(t,\omega) \in [0, +\ns) \times \Omega} Y_t^*(\omega)   <+\ns.
\]
Similarly, a real random variable $X: \Omega \ra \R$ is \emph{globally bounded} if 
$\sup_{\omega \in \Omega} |X(\omega)|  <+\ns$.

Throughout the whole paper we apply the following convention. A sequence of real numbers $d_{n}$, where $n=0,1,2,\ldots$,
is denoted by $\rbr{d_{n}}$ or $\rbr{d_{n}}_{n}$ and a sequence
of real numbers $d^{n}$, where $n=0,1,2,\ldots$, is denoted by $\rbr{d^{n}}$
or $\rbr{d^{n}}_{n}$ (without indication that $n$ ranges over the
set of nonnegative integers $\N$). A similar convention will be applied
to infinite sequences of stopping times, variables etc. 

An $\mathbb{F}$-\emph{stopping time} (or \emph{stopping time} in short) is a random variable $\tau: \Omega \ra [0, +\ns]$ such that for any $t \ge 0$, 
\[
\cbr{\tau \le t}:= \cbr{\omega \in \Omega: \tau(\omega) \le t} \in {\cal F}_t.
\]
A $\sigma$-field ${\cal F}_{\tau}$ generated by the stopping time $\tau$ consists of those events $A \in {\cal F}$ which for any $t \ge 0$ satisfy 
\[
A \cap \cbr{\tau \le t} \in {\cal F}_t.
\]

Since almost all reasonings in this article are pathwise, we will often
omit the argument $\omega\in\Omega$ in formulas, even if the quantities
appearing in these formulas depend on it. 

Now let us introduce sequences of stopping times which we will work
with.

A sequence of $\mathbb{F}$-stopping times $\rbr{\tau_{n}}$ is
called \emph{non-decreasing} if for all $n\in\N$ and \emph{each}
$\omega\in\Omega$, $\tau_{n+1}(\omega)\ge\tau_{n}(\omega)$. 

A sequence of $\mathbb{F}$-stopping times $\rbr{\tau_{n}}$ is
called \emph{proper} if it is non-decreasing, $\tau_{0}\equiv0$ and
for \emph{each} $\omega\in\Omega$ the sequence $\rbr{\tau_{n}(\omega)}$
is divergent to $+\ns$ or there exists some $n\in\N$ such that $\tau_{n}(\omega)=\tau_{n+1}(\omega)=\ldots\in[0,+\ns]$. 

A \emph{simple trading strategy} is a triplet $\rbr{c,\rbr{\tau_{n}},\rbr{g_{n}}}$
which consists of the initial capital $c\in\R$, a proper sequence
of $\mathbb{F}$-stopping times $\rbr{\tau_{n}}$ and a sequence
of globally bounded, real variables $g_{n}:\Omega\ra\R$, $n=0,1,\ldots$, such
that $g_n$ is $\rbr{{\cal F}_{\tau_{n}},{\cal B}\rbr{\R}}$-measurable  and $g_{n}(\omega)=0$ whenever $\tau_{n}(\omega)=+\ns$.

A \emph{step process} $G$ is a real process which may be represented as
$$G_t (\omega)= \sum_{n=1}^{+\ns} g_{n-1}(\omega) {\bf 1}_{\left[\tau_{n-1}(\omega), \tau_n(\omega) \right)} (t)$$
where $\rbr{c,\rbr{\tau_{n}},\rbr{g_{n}}}$ is a simple trading
strategy.

For a real process $X:[0,+\ns)\times\Omega\ra\R$ and a simple trading
strategy $G=\rbr{c,\rbr{\tau_{n}},\rbr{g_{n}}}$ we define 
\[
(G\cdot X)_{t}(\omega):=c+\sum_{n=1}^{+\ns}g_{n-1}(\omega)\rbr{X_{\tau_{n}(\omega)\wedge t}(\omega)-X_{\tau_{n-1}(\omega)\wedge t}(\omega)}.
\]
(For two numbers $a,b\in[-\ns,+\ns]$ we define $a\wedge b=\min\cbr{a,b}$.) Let us note that since the sequence $\rbr{\tau_n}$ is proper, there is only  a finite number of non-zero summands in the sum appearing in the definition of $(G\cdot X)_{t}(\omega)$.

We define the \emph{simple capital process or simple integral} corresponding
to the vector ${\bf G} = \rbr{G^j}_{j \in J}$ of simple trading strategies $G^{j}$, $j\in J$, as 
\[
({\bf G} \cdot {\bf S})_{t}(\omega):=\sum_{j\in J}(G^{j}\cdot S^{j})_{t}(\omega).
\]
The simple capital process has a very natural interpretation -- it is the wealth accumulated till time $t$ by the application
of the simple trading strategy $G^j$ to the asset whose price is represented by the basic martingale $S^j$, $j \in J$. 

\begin{remark} If $G=\rbr{c,\rbr{\tau_{n}},\rbr{g_{n}}}$ is a trading
strategy and for some $\omega\in\Omega$ and $n\in\N$, $\tau_{n}(\omega)=\tau_{n+1}(\omega)=\ldots\in[0,+\ns)$
then the process $G \cdot X$ for the strategy $G$ and a real process $X$, 
is the same for $t\ge\tau_{n}(\omega)$ as if the trading was ceased at $\tau_{n}(\omega)$, even
though $g_{n}(\omega)\neq0$. Thus for all trading strategies we could
add requirement that $g_{n}(\omega)=0$ if $\tau_{n+1}(\omega)=\tau_{n}(\omega) < +\ns$
for some $n\in\N$ (it is possible to verify this condition at the
moment $\tau_{n}(\omega)$), or for a given trading strategy always
modify it so that it satisfies this condition.

Requirement on the sequence $\rbr{\tau^{n}}$ in the
definition of a simple trading strategy to be proper together with
the condition $g_{n}(\omega)=0$ if $\tau_{n+1}(\omega)=\tau_{n}(\omega)$
for some $n\in\N$ guarantees that the trading never occurs with infinite
frequency till any finite time. However, to avoid dealing with
too many technical details we do not add this requirement. 
\end{remark} 

\section{Nonnegative supermartingales, instantly enforceable properties, an outer measure of properties related to the instant enforcement, martingales} \label{sect_defi}
\begin{definition} \label{nsu_defi}
The class ${\cal C}$ of \emph{nonnegative supermartingales} is defined
as the smallest class with the following properties
\begin{enumerate}
\item
${\cal C}$ contains all simple capital processes which are nonnegative;
\item 
whenever $X \in {\cal C}$, $Y$ is a simple capital process and $X + Y$ is nonnegative then $X+Y \in {\cal C}$;
\item
for any sequence $\rbr{X^{n}}$
such that $X^{n}\in{\cal C}$ for $n\in\N$, we have that $X:=\liminf_{n\ra+\ns}X^{n}$
also belongs to ${\cal C}$. 
\end{enumerate}
\end{definition}

Using transfinite induction on the countable ordinals $\alpha$ one may prove that ${\cal C}$ is a convex cone, which means that whenever $X,Z \in {\cal C}$ then for any $s>0$, $s X\in {\cal C}$ and $X+ Z \in {\cal C}$. An elementary reference on the Transfinite Induction Principle is for example \cite[Chapt. 6]{HrbacekJech:1999}.
Indeed, let ${\cal C}^0$ be the class of all simple capital processes which are nonnegative and for $\alpha >0$, $X \in {\cal C}^{\alpha}$ if and only if there exists  $\tilde{X} \in {\cal C}^{<\alpha} : = \bigcup_{\beta < \alpha} {\cal C}^{\beta}$ and a simple capital process  $Y$ such that $X  = \tilde{X}+ Y$ is nonnegative or there exists a sequence of nonnegative supermartingales $X^1, X^2, \ldots$ from ${\cal C}^{<\alpha}$ such that $X=\liminf_{n\ra+\ns}X^{n}$. Using conditions 1. and 2. of Definition \ref{nsu_defi} we have that whenever $X \in {\cal C}^{0}$ and $Z \in {\cal C}$ then for any $s>0$, $s X\in {\cal C}$ and $X+ Z \in {\cal C}$. Assume that the statement 'whenever $X \in {\cal C}^{<\alpha}$ and $Z \in {\cal C}$ then for any $s>0$, $s X\in {\cal C}$ and $X+ Z \in {\cal C}$' holds. Assume now that $X\in {\cal C}^{\alpha}$ and $Z \in {\cal C}$. Considering two possible cases (either $X  = \tilde{X}+ Y$,  $\tilde{X} \in {\cal C}^{<\alpha}$, $Y$ is a simple capital process and $X$ is nonnegative, or $X=\liminf_{n\ra+\ns}X^{n}$, where $X^1, X^2, \ldots \in {\cal C}^{<\alpha}$) we easily get that   $sX  \in {\cal C}$ and $X + Z \in {\cal C}$.  

\begin{remark}
Our definition of the family of nonnegative supermartingales differs slightly from that proposed by Shafer and Vovk, who do not assume the second condition, only the first and the third ones, see \cite[Sect. 14.1]{ShaferVovk:2018}. We need the second condition to prove Fact \ref{conservatism_e_bar}. It remains an open question whether the family $\cal C$ coincides with the family $\cal \tilde{C}$ of nonnegative supermartingales in the sense of Shafer and Vovk.
\end{remark}

In \cite[Sect. 14.1]{ShaferVovk:2018} there is defined a notion of \emph{instant enforcement}
of a subset $E\subseteq[0,+\ns)\times\Omega$ (also called a property
of $t$ and $\omega$). Informally, property $E$ is instantly enforceable
if there exists a trading strategy making a trader using this strategy
infinitely rich as soon as the property $E$ ceases to hold. A formal
definition is the following: a property $E\subseteq[0,+\ns)\times\Omega$
is \emph{instantly enforceable}, or holds \emph{with instant enforcement}, w.i.e. in short, if there exists $X \in {\cal C}$ such that $X_{0}=1$ and 
\[
(t,\omega)\notin E\Longrightarrow X_{t}(\omega)=+\ns.
\]
Complements of instantly enforceable properties (sets) are called
\emph{instantly blockable}. 

The main result of this section is that it is possible to introduce
an outer measure $\overline{\P}$ on the subsets of $[0,+\ns)\times\Omega$
(similarly as in the case of the notion of \emph{null events}, where
one can introduce Vovk's outer probability on all subsets of $\Omega$, cf. \cite{Vovk_volatility:2008}) such that $B\subseteq[0,+\ns)\times\Omega$
is instantly blockable iff $\overline{\P}(B)=0$. 

For $A\subseteq[0,+\ns)\times\Omega$ we define 
\begin{align*}
 \overline{\P}(A) := \inf \cbr{ X_0: X\in{\cal C} \text{ and } \forall(t,\omega)\in[0,+\ns)\times\Omega, X_{t}(\omega)\ge{\bf 1}_{A}(t,\omega) }.
\end{align*}
We have the following lemma. 
\begin{lemma}  \label{mainlema}
The set $B\subseteq[0,+\ns)\times\Omega$ is instantly
blockable iff
\[
\overline{\P}(B)=0.
\]
\end{lemma}
\begin{proof} If $B$ is instantly blockable then there exists $X\in{\cal C}$
such that $X_{0}=1$ and 
\[
(t,\omega)\in B\Longrightarrow X_{t}(\omega)=+\ns.
\]
Thus, taking arbitrary $\varepsilon>0$ we have $\rbr{\varepsilon X}_{0}=\varepsilon$
and 
\[
(t,\omega)\in B\Longrightarrow\rbr{\varepsilon X}_{t}(\omega)=+\ns>{\bf 1}_{B}(t,\omega);
\]
\[
(t,\omega)\notin B\Longrightarrow\rbr{\varepsilon X}_{t}(\omega)\ge0={\bf 1}_{B}(t,\omega);
\]
and since $\varepsilon X\in{\cal C}$ we get 
\[
\overline{\P}(B)\le\varepsilon.
\]
Since $\varepsilon$ may be as close to $0$ as we wish, $\overline{\P}(B) \le 0$ and thus $\overline{\P}(B) = 0$ 
(the opposite inequality $\overline{\P}(B)\ge0$ holds since for any
$X\in{\cal C}$, $X_{0}\ge0$). 

Assume now that $\overline{\P}(B)=0$. For $n=1,2,\ldots$, there
exists $X^{n}\in{\cal C}$ such that $X_{0}^{n}(\omega)\le2^{-n}$
and for all $(t,\omega)\in[0,+\ns)\times\Omega$,
\[
(t,\omega)\in B\Longrightarrow X_{t}^{n}(\omega)\ge{\bf 1}_{B}(t,\omega)=1.
\]
Taking $X=\rbr{1-\sum_{n=1}^{+\ns}X_{0}^{n}}+\sum_{n=1}^{+\ns}X^{n}$
we get $X\in{\cal C}$, $X_{0}=1$ (notice that $\sum_{n=1}^{+\ns}X_{0}^{n}\le1$)
and 
\[
(t,\omega)\in B\Longrightarrow X_{t}(\omega)\ge\sum_{n=1}^{+\ns}X_{t}^{n}(\omega)\ge\sum_{n=1}^{+\ns}1=+\ns.
\]
\end{proof}

\begin{remark}
Replacing in the definition of {instantly enforceable} property and in the definition of $\overline{\P}$ the family $\cal C$ by $\cal \tilde{C}$ (the family of nonnegative supermartingales in the sense of Shafer and Vovk) one obtains an outer measure which assigns zero value exactly to those sets (properties) of pairs of time $t$ and an elementary event $\omega$ which are instantly blockable in the sense of Shafer and Vovk. The proof is exactly the same as the proof of Lemma \ref{mainlema}.
\end{remark}

The definition of upper probability $\overline{\P}$ may be generalized
and we may define the \emph{upper expectation} (or cost of super-hedging
or super-replication) of a generalized process $Y:[0,+\ns)\times\Omega\ra[-\ns,+\ns]$
in the following way
\begin{align*}
\overline{\E}Y := \inf \cbr{ X_0: X\in{\cal C} \text{ and } \forall(t,\omega)\in[0,+\ns)\times\Omega, X_{t}(\omega)\ge Y_{t}(\omega)}.
\end{align*}

For $A\subseteq[0,+\ns)\times\Omega$ we have 
\[
\overline{\P}(A)=\overline{\E}{\bf 1}_{A}.
\]

For two generalized processes $X$ and $Y$ we say that $X$
\emph{dominates} $Y$ if they satisfy the condition
\[
\forall(t,\omega)\in[0,+\ns)\times\Omega,\quad X_{t}(\omega)\ge Y_{t}(\omega).
\]
For two generalized processes $X$ and $Y$ we say that $X$
\emph{dominates $Y$ with instant enforcement (w.i.e.)}  if the set of $(t, \omega)$ where the inequality
$X_{t}(\omega)\ge Y_{t}(\omega)$
holds is instantly enforceable. 

Below we list, without proofs, properties of $\overline{\E}$ which imply that
$\overline{\P}$ is an outer measure:

1) non-negativity: for any generalized process $Y$, $\overline{\E}Y\ge0$;

2) monotonicity with respect to domination of generalized processes: if $Z$ dominates $Y$ or $Z$ dominates $Y$ w.i.e. and $Z_t(\omega) > -\ns$ w.i.e. then 
\[
\overline{\E}Y\le\overline{\E}Z;
\]

3) positive homogeneity: if $\alpha\in [0, +\ns)$ then 
\[
\overline{\E}(\alpha Y)=\alpha\overline{\E}Y,
\]
where we apply the convention that $0 \cdot (\pm\ns) = 0$;

4) countable subadditivity for nonnegative generalized processes: if  $\forall(t,\omega)\in[0,+\ns)\times\Omega$, $Y_{t}^{1}(\omega),Y_{t}^{2}(\omega),\ldots\ge0$ 
then
\[
\overline{\E}\rbr{\sum_{k=1}^{+\ns}Y^{k}}\le\sum_{k=1}^{+\ns}\overline{\E}Y^{k}.
\]

5) finite subadditivity for generalized processes not attaining $-\ns$ (this is assumed in order to be able to calculate $\sum_{m=1}^{n}Y^{m}$):  if $Y^m: [0,+\ns)\times\Omega \ra (-\ns, +\ns]$, $m=1,2,\ldots,n$ ($n \in \N$), are generalized processes 
then
\[
\overline{\E}\rbr{\sum_{m=1}^{n}Y^{m}}\le\sum_{m=1}^{n}\overline{\E}Y^{m};
\]

6) consistency: if $Y_t(\omega) = 1$ w.i.e. then $\overline{\E}Y = 1$;

7) a lower bound for generalized processes: for any $Y:[0,+\ns)\times\Omega$, $\overline{\E}Y \ge \max\cbr{Y_0,0}$.

8) determinism for nonnegative supermartingales: if $Y$ is a nonnegative supermartingale then $\overline{\E}Y = Y_0$.

Also, an almost immediate consequence of the definition of $\overline{\E}$
is the Fatou lemma.

\begin{fact}[Fatou's lemma] \label{98} If $\rbr{Y^{n}}$ is a sequence of
generalized processes then 
\[
\overline{\E}\liminf_{n\ra+\ns}Y^{n}\le\liminf_{n\ra+\ns}\overline{\E}Y^{n}.
\]
\end{fact}
\begin{proof} Let $X^{n}\in{\cal C}$, $n=1,2,\ldots$, be such that $X_{0}^{n}\le\overline{\E}Y^{n}+1/n$
and $\forall(t,\omega)\in[0,+\ns)\times\Omega$, $X_{t}^{n}(\omega)\ge Y_{t}^{n}(\omega)$.
Denote $X=\liminf_{n\rightarrow+\ns}X^{n}$ then 
\[
\forall(t,\omega)\in[0,+\ns)\times\Omega,\quad X_{t}(\omega)=\liminf_{n\rightarrow+\ns}X_{t}^{n}(\omega)\ge\liminf_{n\rightarrow+\ns}Y_t^{n}(\omega).
\]
Hence, since $X\in{\cal C}$, 
\[
\overline{\E}\liminf_{n\ra+\ns}Y^{n}\le X_{0}=\liminf_{n\rightarrow+\ns}X_{0}^{n}\le\liminf_{n\rightarrow+\ns}\overline{\E}Y^{n}.
\]
\end{proof}
\begin{definition} \label{cont_defi}
A process $Y$ has trajectories which are continuous w.i.e. ($Y$ is continuous w.i.e. in short) if the set of pairs $(t, \omega) \in [0, +\ns) \times \Omega$ such that the trajectory $[0, +\ns) \ni s \mapsto Y_s(\omega)$ is real and continuous at the point $t$ is instantly enforceable.
\end{definition}
Now we will prove a technical lemma which we will use to prove that some quantities we define in the sequel are stopping times. 
\begin{lemma} \label{stop_times}
Let $Y$ be a process whose trajectories are continuous w.i.e. and $\mathbb{F} = \rbr{{\cal F}_t}_{t \ge 0}$ be a filtration satisfying Assumption {\bf A}. Let $\tau: \Omega \ra [0, +\ns]$ be a stopping time (with respect to the filtration $\mathbb{F}$) and let $F$ be a closed subset of $\R$. Define  $\rho: \Omega \ra [0, +\ns]$ by 
\[
\rho(\omega) =
\inf \cbr{ t \ge \tau(\omega): Y_t \in F } 
\]
with the usual convention that $\inf \emptyset = +\ns$.
Then $\rho$ is a stopping time with respect to the filtration $\mathbb{F}$.
If moreover $\tau$ is such that for each $\omega \in \Omega \cap \cbr{\tau < +\ns}$, $Y_{\tau(\omega)}(\omega)$ is an isolated point of $F$ then $\sigma: \Omega \ra [0, +\ns]$ defined by 
\[
\sigma(\omega) =
\inf \cbr{ t > \tau(\omega): Y_t \in F \setminus \cbr{Y_{\tau(\omega)}} } 
\]
is also  a stopping time with respect to the filtration $\mathbb{F}$.
\end{lemma} 
\begin{proof}
Let us fix $t \in [0, +\ns)$. We need to prove that $\cbr{\rho \le t}, \cbr{\sigma \le t} \in {\cal F}_t$. Let $C \subseteq [0, +\ns) \times \Omega$ be the set of pairs $(s, \omega) \in [0, +\ns) \times \Omega$ such that the trajectory $[0, +\ns) \ni s \mapsto Y_s(\omega)$ is continuous at the point $s$. By assumption, its complement $B = \rbr{[0,+\ns) \times \Omega} \setminus C$ is instantly blockable and thus, by Assumption {\bf A},  the projection of $B \cap \rbr{[0,t] \times \Omega}$ onto $\Omega$ belongs to ${\cal F}_{t}$. Let us denote this projection by $\Omega_B$. For $y \in \R$ let $d\rbr{y, F} = \inf\cbr{|y-z|: z \in F}$ denote the distance of $y$ from the set $F$ and let $\mathbb{Q}$ denote the set of all rational numbers.
For each $\omega \in \Omega \setminus \Omega_B$ the trajectory $[0, t] \ni s \mapsto Y_s(\omega)$ is continuous (if  $[0, t] \ni s \mapsto Y_s(\omega)$ was not continuous at some point $s\in [0,t]$ then $(s, \omega) \in B$ and thus $\omega \in \Omega_B$) and we get
\begin{align*}
& \cbr{\omega \in  \Omega \setminus \Omega_B: \rho(\omega) \le t} = \cbr{\omega \in  \Omega \setminus \Omega_B:  \inf_{ \tau(\omega) \le s \le t, s \in \mathbb{Q} \cup \cbr{t}} d \rbr{Y_s(\omega), F} =0 } \\
&  = \bigcap_{n \in \N} \bigcup_{0 \le s \le t, s \in \mathbb{Q} \cup \cbr{t}} \cbr{\omega \in  \rbr{\Omega \setminus \Omega_B} \cap \cbr{\tau \le s}:  d\rbr{Y_s(\omega), F}  \le \frac{1}{n+1} } \in {\cal F}_t.
\end{align*}
Next we consider
$
\cbr{\omega \in  \Omega_B: \rho(\omega) \le t} = \Omega_B \cap \cbr{\rho \le t}.
$
There exists a subset $D$ of $B$ whose projection equals $\Omega_B \cap \cbr{\rho \le t}$. $D$ being a subset of $B$ is instantly blockable and its projection onto $\Omega$ belongs to ${\cal F}_t$. 
Thus, 
\[
\cbr{\rho \le t} = \cbr{\omega \in  \Omega \setminus \Omega_B: \rho(\omega) \le t} \cup \cbr{\omega \in  \Omega_B: \rho(\omega) \le t} \in {\cal F}_t.
\] 

To prove that $\cbr{\sigma \le t} \in {\cal F}_t$ we write 
\begin{align*}
& \cbr{\omega \in  \Omega \setminus \Omega_B: \sigma(\omega) \le t} \\& = \cbr{\omega \in  \Omega \setminus \Omega_B:  \inf_{\tau(\omega) < s \le t, s \in \mathbb{Q}\cup \cbr{t} } d\rbr{Y_s(\omega), F\setminus \cbr{Y_{\tau(\omega)}}} =0 } \in {\cal F}_t.
\end{align*}
The rest of the proof is the same as for $\rho$.
\end{proof}

Next to the class of nonnegative supermartingales, other important
class of processes which we will work with is the family of \emph{martingales}.
The class of martingales ${\cal M}$ is defined as the smallest $\lim$-closed
class of real (w.i.e.) processes such than it contains all simple capital processes.
By the fact that ${\cal M}$ is $\lim$-closed we mean that whenever
$X^{n}\in{\cal M}$, $n\in\N$, and $X$ is a real (w.i.e.) process such that for any $(t,\omega)\in[0,+\ns)\times\Omega$,
\begin{equation}
\lim_{n\ra+\ns}\sup_{s\in[0,t]}\left|X_{s}(\omega)-X_{s}^{n}(\omega)\right|=0\quad{ w.i.e.}\label{eq:unifconv}
\end{equation}
then also $X\in{\cal M}$. 
\begin{remark}
To deal with the improper (or non-existent) limits of sequences of processes, Shafer and Vovk introduce in  \cite[Sect. 14.1]{ShaferVovk:2018} also a 'cemetery' state $\partial$, which may be attained by martingales  since some moment in time, but in this article we will deal with martingales attaining values in  $[-\ns, +\ns]$ only.
\end{remark}
Using condition (\ref{eq:unifconv}) and transfinite induction we get the following fact
\begin{fact} \label{cont_of_mart}
Let $X$ be a martingale. The property of $\rbr{t, \omega} \in [0, +\ns)\times \Omega$ that the trajectory $[0, +\ns) \ni s \mapsto X_s(\omega)$ is real and continuous on $[0, t]$ is instantly enforceable.
\end{fact}
\begin{proof}
We will use transfinite induction on the countable ordinals $\alpha$. 
Let ${\cal M}^0$ be the class of all simple capital processes and for $\alpha >0$, $X \in {\cal M}^{\alpha}$ if and only if there exists a sequence $X^1, X^2, \ldots$ of martingales in ${\cal M}^{<\alpha} = \bigcup_{\beta < \alpha} {\cal M}^{\beta}$ such that \eqref{eq:unifconv} holds.

Naturally, all martingales in ${\cal M}^{0}$ are continuous. Assume that for a given ordinal $\alpha$, for all martingales in ${\cal M}^{<\alpha}$ the trajectories $[0, +\ns) \ni s \mapsto X_t^n(\omega)$ are real and continuous on $[0, t]$ w.i.e. Now let $X \in {\cal M}^{\alpha}$ and let $\rbr{X^n}$ be a sequence of martingales in ${\cal M}^{<\alpha} $ such that \eqref{eq:unifconv} holds. Let $E$ be the intersection of the sets of pairs $\rbr{t, \omega} \in [0, +\ns)\times \Omega$ such that the trajectories $[0, +\ns) \ni s \mapsto X_t^n(\omega)$ are real and continuous on $[0, t]$ and where \eqref{eq:unifconv} holds. $E$, as the intersection of countably many instantly enforceable sets, is instantly enforceable. From  \eqref{eq:unifconv}  it follows that for $\rbr{t, \omega} \in E$ the trajectory $[0, +\ns) \ni s \mapsto X_s(\omega)$ is real and continuous on $[0, t]$, which finishes the proof.
\end{proof}
\begin{remark}
In the proof of Fact  \ref{cont_of_mart} we also quietly used the fact that using the definition of $\cal M$ and starting from ${\cal M}^{0}$ (the family of all simple capital processes) we recover whole $\cal M$, that is we do not need to 'produce' more martingales than those which are in $\bigcup_{\alpha} {\cal M}^{\alpha}$, where $\alpha$ ranges over all countable ordinals ($\cal M$ is supposed to be minimal). 
   
   A more precise reasoning is the following. Denote ${\cal M}^{\cup}:=\bigcup_{\alpha} {\cal M}^{\alpha}$, where $\alpha$ ranges over all countable ordinals.  To prove that ${\cal M}^{\cup} = {\cal M}$ we need to prove that 
   \begin{enumerate}
\item ${\cal M}^{0} \subseteq {\cal M}^{\cup}$;  
\item whenever $X^{n}\in{\cal M}^{\cup}$, $n\in\N$, and $X$ is a real (w.i.e.) process such that 
\begin{equation*}
\lim_{n\ra+\ns}\sup_{s\in[0,t]}\left|X_{s}(\omega)-X_{s}^{n}(\omega)\right|=0\quad{ w.i.e.}\label{eq:unifconv}
\end{equation*}
then also $X\in{\cal M}^{\cup}$. 
\end{enumerate}
The fact that ${\cal M}^{0} \subseteq {\cal M}^{\cup}$ is trivial. 
  To prove the second fact, assume that $X^n \in {\cal M}^{\alpha_n}$, where $\alpha_n$ is a countable ordinal. Let $\alpha_{\ns}$ be the smallest ordinal greater than all $\alpha_n$, $n \in \N$. $\alpha_{\ns}$  is also countable and by the definition of the sets ${\cal M}^{\alpha}$, $X \in {\cal M}^{\alpha_{\ns}}$, thus also $X \in {\cal M}^{\cup}$. 
\end{remark}
We have the following important fact. 
\begin{fact} \label{conservatism_e_bar}
Assume that $Y$ is a nonnegative supermartingale and $X$ is a martingale such that $Y+X$ is nonnegative, then
 there exists a nonnegative supermartingale $Z$ such that $Y + X = Z$ w.i.e.
\end{fact}
\begin{proof}
Again, we will use transfinite induction on the countable ordinals $\alpha$. We apply the same notation as in the proof of Fact \ref{cont_of_mart}. 
If $X \in {\cal M}^0$ then $Y + X$ is also a nonnegative supermartingale (condition 2. in Definition \ref{nsu_defi}). 
 Let now $X \in {\cal M}^{\alpha}$ for some $\alpha >0$ and let $X^1, X^2, \ldots$ be martingales in ${\cal M}^{<\alpha}$ such that \eqref{eq:unifconv} holds. For fixed $\varepsilon >0$ we consider the following processes:
\begin{align}
A^{\varepsilon, n}_t & := Y_{t\wedge  \sigma^{\varepsilon, n}}  + \varepsilon   + X^n_{t\wedge  \sigma^{\varepsilon, n}}, 
\end{align}
 where 
\[
\sigma^{\varepsilon, n} := \inf\cbr{s >0: X^n_s - X_s \le  - \varepsilon}.
\]
From Lemma \ref{stop_times} (and Fact \ref{cont_of_mart}) it follows that $\sigma^{\varepsilon, n}$ is a stopping time and it is easy to see that $Y_{t\wedge  \sigma^{\varepsilon, n}}$ is a nonnegative supermartingale while $\varepsilon   + X^n_{t\wedge  \sigma^{\varepsilon, n}}$ is a martingale from ${\cal M}^{<\alpha}$. Moreover, $A^{\varepsilon, n}$ is nonnegative (it follows from the fact that $Y+X$ is nonnegative and from the definition of $\sigma^{\varepsilon, n}$), thus, by the inductive assumption, $A^{\varepsilon, n}$ is equal to a nonnegative supermartingale $Z^{\varepsilon, n}$ w.i.e. 
We notice that $Z^{\varepsilon} := \liminf_{n \ra +\ns} Z^{\varepsilon, n} = \varepsilon + Y + X$ on the set $\Omega^{\varepsilon}$ where \eqref{eq:unifconv} holds as well all the equalities $A^{\varepsilon, n} = Z^{\varepsilon, n}$ hold. 

Choosing a sequence $\rbr{\varepsilon_m}$ such that $\varepsilon_m >0$ and $\varepsilon_m \ra 0$ as $m \ra +\ns$ we get  that 
$ Y + X = Z := \liminf_{m \ra +\ns} Z^{\varepsilon_m} $ on the set $\bigcap_{m} \Omega^{\varepsilon_m}$. Thus $X+Y = Z\in {\cal C}$ w.i.e.
\end{proof} 

\begin{corollary} \label{ebar_est}
If $Y\in {\cal C}$, $X \in {\cal M}$ is such that $X_0 =0$ and $Y+X$ is nonnegative then $\overline{\E} \rbr{Y +X} = Y_0$.
\end{corollary}
\begin{proof}
By Fact \ref{conservatism_e_bar} there exists $Z \in {\cal C}$ such that $Y+X = Z$ w.i.e.  
Let us fix $\varepsilon>0$ and let $U\in {\cal C}$ be such that $U_0 \le \varepsilon$ and $U= +\ns$ on the set where $X+Y \neq Z$. We have that $Z+U \in {\cal C}$ dominates $Y+X$  hence
\[
Y_0 = Y_0 +X_0 \le \overline{\E} \rbr{Y +X} \le \overline{\E} \rbr{Z +U}  = Z_0 + U_0 \le Z_0 + \varepsilon. 
\]
On the other side, since $Y = Z-X$ w.i.e and $U= +\ns$ on the set where $Y \neq Z-X$, $Y+U$ dominates $Z-X$ and hence 
\[
Z_0 = Z_0 - X_0 \le \overline{\E} \rbr{Z-X} \le \overline{\E} \rbr{Y +U}  = Y_0 + U_0 \le Y_0 + \varepsilon. 
\]
Since $\varepsilon$ is an arbitrary positive real we must have  $\overline{\E} \rbr{Y +X} = Y_0 = Z_0$.
\end{proof} 
\begin{remark} \label{conservatism}
By transfinite induction it is possible to prove (see \cite[Sect. 14.2]{ShaferVovk:2018}) that whenever $X$ is a martingale and $G$ is a simple trading strategy then $G\cdot X$ is again a martingale. We will use this in the sequel.
\end{remark}

\section{Simple quadratic variation - definition, It\^o's isometry and BDG inequalities}

Let $X$ be a martingale and $\tau=\rbr{\tau_{n}}$ be a proper sequence
of stopping times. We define the \emph{simple quadratic variation process
of $X$ along $\tau$} as 
\[
\sbr X_{t}^{\tau}:=\sum_{n=1}^{+\ns}\rbr{X_{\tau_{n}\wedge t}-X_{\tau_{n-1}\wedge t}}^{2},\quad t\in[0,+\ns).
\]

\begin{lemma} \label{100} Let $X$ be a martingale and $\tau=\rbr{\tau_{n}}$
be a proper sequence of stopping times. The process 
\[
Y_{t}:=\rbr{X_{t}-X_{0}}^{2}-\sbr X_{t}^{\tau},\quad t\in[0,+\ns),
\]
is a martingale. 
\end{lemma}

\begin{proof} For $M\in(0,+\ns)$ let $\sigma(M)=\sigma(X,M)$ denote
the stopping time defined as
\begin{equation}
\sigma(X,M):=\inf\cbr{t\in[0,+\ns):\left|X_{t}\right|\ge M}.\label{eq:sigmaM}
\end{equation}
By Lemma \ref{stop_times}, $\sigma(X,M)$ is indeed a stopping time.
 Now let us define the simple trading strategy $G^{M}=\rbr{0,\rbr{\tau_{n}},\rbr{g_{n}^{M}}}$
with $g_{0}:\R\ra\R$, $g_{0}(\omega):=0$ and $g_{n}:\R\ra\R$,
\begin{align*}
g_{n}^{M}(\omega): & =\begin{cases}
2\rbr{X_{\tau_{n}}(\omega)-X_{0}(\omega)} & \text{if }n\in\N\text{ and }\tau_{n}<\sigma(M),\\
0 & \text{if }n\in\N\text{ and }\tau_{n}\ge\sigma(M).
\end{cases}
\end{align*}
All variables $g_{n}$, $n=0,1,2,\ldots$, are bounded, thus $G^M$ is indeed a simple trading strategy.
A direct calculation for $(t, \omega)\in [0, +\ns)\times\Omega$ gives 
\[
Y_{t}^{M}:=Y_{t\wedge\sigma(M)}=(G^{M}\cdot X)_{t}
\]
and thus (recall Remark \ref{conservatism}) $Y^{M}$
is a martingale. Moreover, for $t\in[0,+\ns)$ and a real $M>\sup_{s\in[0,t]}|X_{s}(\omega)|$
we have $Y_{s}^{M}(\omega)=Y_{s}(\omega)$ for $s \in [0, t]$ (we can always find such a $M$ if the trajectory $[0,t] \ni s \mapsto X_s(\omega)$ is real and continuous) so 
\[
\lim_{M\ra+\ns}\sup_{s\in[0,t]}\left|Y_{s}^{M}(\omega)-Y_{s}(\omega)\right|=0 \quad w.i.e.
\]
and hence $Y$ is a martingale.
\end{proof}

\begin{fact}[It\^o's isometry for simple quadratic variation] \label{101} Let $X$ be a martingale and $\tau=\rbr{\tau_{n}}$ be a proper sequence
of stopping times. We have
\[
\overline{\E}\rbr{X-X_{0}}^{2}=\overline{\E}\sbr X^{\tau}.
\]
\end{fact}
\begin{proof} Let $\varepsilon>0$ and $Y$ be a nonnegative supermartingale
such that $Y_{0}\le\overline{\E}\rbr{X-X_{0}}^{2}+\varepsilon$ and
$\forall(t,\omega)\in[0,+\ns)\times\Omega$, $Y_{t}(\omega)\ge\rbr{X_{t}(\omega)-X_{0}(\omega)}^{2}$.
We have 
\[
\forall(t,\omega)\in[0,+\ns)\times\Omega\quad Y_{t}(\omega)-\rbr{X_{t}(\omega)-X_{0}}^{2}+\sbr X_{t}^{\tau}(\omega)\ge\sbr X_{t}^{\tau}(\omega)\ge0.
\]
$Y-\rbr{\rbr{X-X_{0}}^{2}-\sbr X^{\tau}}$ is nonnegative and by Lemma \ref{100}, $\rbr{X-X_{0}}^{2}-\sbr X^{\tau}$ is a martingale (starting from $0$), thus by Corollary \ref{ebar_est}, the following estimates follow
\[
\overline{\E}\sbr X^{\tau}\le \overline{\E} \cbr{Y-\rbr{\rbr{X-X_{0}}^{2}-\sbr X^{\tau}} } = Y_{0}\le\overline{\E}\rbr{X-X_{0}}^{2}+\varepsilon.
\]
Since $\varepsilon$ may be as close to $0$ as we wish, we have 
\[
\overline{\E}\sbr X^{\tau}\le\overline{\E}\rbr{X-X_{0}}^{2}.
\]

The opposite inequality follows by a similar reasoning --  if $Y$ is
a nonnegative supermartingale that dominates $\sbr X^{\tau}$ and
such that $Y_{0}\le\overline{\E}\sbr X^{\tau}+\varepsilon$ then we apply Corollary \ref{ebar_est} to the process $Y-\sbr X^{\tau}+\rbr{X-X_{0}}^{2}$ which dominates $\rbr{X-X_{0}}^{2}$.
\end{proof}
\begin{remark} The proof of Fact \ref{101} may be easily adapted to prove the following, more general statement: if there are two nonnegative, real w.i.e. processes
$X$ and $Y$ whose difference is a martingale starting from $0$ then 
\[
\overline{\E}X=\overline{\E}Y.
\]
\end{remark}
Now we proceed to the Burkholder-Davis-Gundy
inequalities for the simple quadratic variation along some proper sequence of stopping times. As it is one of the main ingredients in the proof of the next fact,
let us briefly recall the pathwise version of the Burkholder-Davis-Gundy
inequalities (BDG inequalities in short) of Beiglboeck and Siorpaes \cite{Beiglboeck:2015}.
Let $x_{k}$, $k\in\N$, be a sequence of real numbers and for $k\in\N$
define
\[
x_{k}^{*}:=\max_{l=0,1,\ldots,k}\left|x_{l}\right|,\quad[x]_{k}:=x_{0}^{2}+\sum_{l=1}^{k}\rbr{x_{l}-x_{l-1}}^{2}
\]
then 
\begin{equation}
x_{k}^{*}\le6\sqrt{[x]_{k}}+2\rbr{h\cdot x}_{k}\text{ and }\sqrt{[x]_{k}}\le3x_{k}^{*}-\rbr{h\cdot x}_{k},\label{eq:BDG1}
\end{equation}
where 
\begin{equation}
\rbr{h\cdot x}_{k}=\sum_{l=1}^{k}h_{l-1}\rbr{x_{l}-x_{l-1}}\text{ with }h_{l}=\frac{x_{l}}{\sqrt{[x]_{l}+x_{l}^{*}}}\label{eq:hdef}
\end{equation}
and we apply the convention that $\frac{0}{0}=0$. Inequalities (\ref{eq:BDG1})
may be viewed as a pathwise version of the BDG inequalities for $p=1$.
To formulate a pathwise version of the BDG inequalities for $p>1$,
for $k,l\in\N$, $k\ge l$, we introduce 
\[
e_{k}^{(l)}:=\frac{x_{k}-x_{l-1}}{\sqrt{[x]_{k}-[x]_{l-1}+\max_{l\le m\le k}\rbr{x_{m}-x_{l-1}}^{2}}},
\]
\[
f_{k}:=p^{2}\sum_{l=0}^{k}\rbr{\sqrt{[x]_{l}^{p-1}}-\sqrt{[x]_{l-1}^{p-1}}}e_{k}^{(l)},
\]
\begin{equation}
g_{k}:=p^{2}\sum_{l=0}^{k}\rbr{\rbr{x_{l}^{*}}^{p-1}-\rbr{x_{l-1}^{*}}^{p-1}}e_{k}^{(l)},\label{eq:fgdef}
\end{equation}
where together with the convention $\frac{0}{0}=0$ we also use $x_{-1}=x_{-1}^{*}=[x]_{-1}=0$.
With the just defined quantities and $\rbr{f\cdot x}_{k}$, $\rbr{g\cdot x}_{k}$
defined similarly as $\rbr{h\cdot x}_{k}$ one has the following pathwise
versions of the BDG inequalities for $p>1$: if $C_{p}=6^{p}(p-1)^{p-1}$
then for $k\in\N$
\begin{equation}
\rbr{x_{k}^{*}}^{p}\le C_{p}\sqrt{[x]_{k}^{p}}+2\rbr{g\cdot x}_{k}\text{ and }\sqrt{[x]_{k}^{p}}\le C_{p}\rbr{x_{k}^{*}}^{p}-\rbr{f\cdot x}_{k}.\label{eq:BDG>1}
\end{equation}

Now, for a generalized process $Y$ and a proper sequence of stopping times
$\tau=\rbr{\tau_{n}}$ we define a process 
\[
Y_{t}^{\tau,*} : =\max_{n\in\N}\left|Y_{\tau_{n}\wedge t}\right|,\quad t\in[0,+\ns)
\]
($\max_{n\in\N}\left|Y_{\tau_{n}\wedge t}\right|$ is
well defined since $\tau$ is proper). 

\begin{fact}[BDG inequalities for simple quadratic variation] \label{99} Let $X$ be a
martingale and $\tau=\rbr{\tau_{n}}$ be a proper sequence of stopping
times. For any $p\ge1$ there exist finite, positive constants $c_{p}$
and $C_{p}$ such that 
\[
c_{p}\overline{\E}\rbr{\sbr X^{\tau}}^{p/2}\le\overline{\E}\rbr{\rbr{X-X_{0}}^{\tau,*}}^{p}\le C_{p}\overline{\E}\rbr{\sbr X^{\tau}}^{p/2}.
\]
In the case $p>1$ one may take $C_{p}=6^{p}(p-1)^{p-1}$ and $c_{p}=1/C_{p}$,
while in the case $p=1$ one may take $C_{p}=6$ and $c_{p}=1/3$.
\end{fact}

\begin{proof} The proof is almost a straightforward application of the pathwise
versions of the BDG inequalities. 

If $p>1$ we fix a real $M>0$, recall
the stopping time $\sigma(M)=\sigma(X,M)$ defined in (\ref{eq:sigmaM})
and define a simple strategy $G^{M}=\rbr{0,\rbr{\tau_{n}},\rbr{g_{n}^{M}}}$
in the following way: for $\omega\in\Omega$ and $n\in\N$ we define
$x_{n}=X_{\tau_{n}}(\omega)-X_{0}$ and
\begin{align*}
g_{n}^{M}(\omega): & =\begin{cases}
g_{n} & \text{if }n\in\N\text{ and }\tau_{n}<\sigma(M),\\
0 & \text{if }n\in\N\text{ and }\tau_{n}\ge\sigma(M),
\end{cases}
\end{align*}
where $g_{n}$ for the given sequence $\rbr{x_{n}}$ is defined as
in (\ref{eq:fgdef}). Functions $g_{n}^{M}$ are globally bounded
(by constants depending on $n$ and $M$) and ${\cal F}_{\tau_{n}}$-measurable.
The pathwise limit $G_{t}^{X}(\omega):=\lim_{M\ra+\ns}\rbr{G^{M}\cdot X}_{t}(\omega)$,
$(t,\omega)\in[0,+\ns)\times\Omega$, is well defined  on the set of $(t, \omega)$ where the trajectory $[0,t] \ni s \mapsto X_s(\omega)$ is real and continuous, since $\rbr{G^{M}\cdot X}_{s}(\omega)$, $s \in [0, t]$,
is the same for all $M>\sup_{s\in[0,t]}\left|X_{s}(\omega)\right|$. For all other $(t, \omega)$ we define $G_{t}^{X}(\omega):=0$ and obtain a martingale $G^{X}$.

Now, by (\ref{eq:BDG>1}), if $Y$ is a nonnegative supermartingale
dominating $\rbr{\sbr X^{\tau}}^{p/2}$ then $C_{p}Y+2G^{X}$ is a
process dominating $\rbr{\rbr{X-X_{0}}^{\tau,*}}^{p}$ w.i.e.
(this follows from the application of (\ref{eq:BDG>1}) to the sequence
$\tilde{x}_{k}=X_{\tau_{k}\wedge t}-X_{0}$, $k\in\N$). By this and Corollary \ref{ebar_est}, similarly as in the proof of It\^o's isometry, we infer
\[
\overline{\E}\rbr{\rbr{X-X_{0}}^{\tau,*}}^{p}\le C_{p}\overline{\E}\rbr{\sbr X^{\tau}}^{p/2}.
\]
The inequality 
\[
\overline{\E}\rbr{\rbr{X-X_{0}}^{\tau,*}}^{p}\ge c_{p}\overline{\E}\rbr{\sbr X^{\tau}}^{p/2}
\]
with $c_{p}=1/C_{p}$ may be proven similarly, with the help of the sequence
$\rbr{f_{n}}$. 

The case $p=1$ is even easier since one does not
need to use the stopping time $\sigma(M)$ to define appropriate trading
strategies, since $h_{l}$, $l\in\N$, in (\ref{eq:BDG1}) always
belong to the interval $[-1,1]$.
\end{proof}

\section{Quadratic variation -- existence, It\^o's isometry and BDG inequalities}
\subsection{Quadratic variation -- existence}
In this section we will prove that the simple quadratic variations of a martingale along sequences of stopping times satisfying some condition converge w.i.e. To formulate this condition we need to define a fine cover of a real
process. 

A non-decreasing sequence of $\mathbb{F}$-stopping
times $\rbr{\tau_{n}}$ is called \emph{a fine cover} \emph{of the
process $X$ with accuracy }$\delta>0$ (or: $\rbr{\tau_{n}}$
\emph{finely covers} \emph{the process $X$ with accuracy }$\delta>0$) \emph{on the set} $E \subseteq [0, +\ns) \times \Omega$
if $\tau_{0}\equiv0$, for any $(t, \omega) \in E$ there are only finitely many $n \in \N$ such that $\tau_{n}(\omega) \le t$ and for any $n\in \N$ and $(t, \omega) \in E$ 
\begin{equation} \label{fin_cov_wie}
\sup_{s\in\sbr{\tau_{n}(\omega) \wedge t,\tau_{n+1}(\omega)\wedge t}}X_{s}-\inf_{s\in\sbr{\tau_{n}(\omega) \wedge t,\tau_{n+1}(\omega)\wedge t}}X_{s}\le\delta.
\end{equation}
If the set $E$ is instantly enforceable then we say that the sequence $\rbr{\tau_{n}}$ is \emph{a fine cover} \emph{of the
process $X$ with accuracy }$\delta>0$ (or: $\rbr{\tau_{n}}$
\emph{finely covers} \emph{the real process $X$ with accuracy }$\delta>0$) \emph{ w.i.e.}

Using ideas from \cite{ShaferVovk:2018}, which may be attributed already to Kolmogorov,  we first prove the following lemma.

\begin{lemma} \label{102} Let $X$ be a martingale, $\sigma=\rbr{\sigma_{n}}$
be a fine cover of $X$ with accuracy $\delta>0$ on the set $E \subseteq [0, +\ns) \times \Omega$, $\tau=\rbr{\tau_{n}}$
be a sequence of $\mathbb{F}$-stopping times such that for any $(t, \omega) \in E$ there are only finitely many $n \in \N$ such that $\tau_{n}(\omega) \le t$, and let $\upsilon$ be the non-decreasing rearrangement
of the stopping times from both sequences $\sigma$ and $\tau$, $\upsilon=\rbr{\upsilon_{n}}$,
with redundancies deleted. Then 
\begin{equation}
\sbr{\sbr X^{\sigma}-\sbr X^{\upsilon}}^{\upsilon}\le4\delta^{2}\sbr X^{\upsilon} \text{ on } E\label{eq:q_var_estim}
\end{equation}
\end{lemma}
\begin{proof} Let $(t,\omega) \in E$. In all formulas which follow in the proof we omit $\omega$. We have 
\[
\sbr{\sbr X^{\sigma}-\sbr X^{\upsilon}}_{t}^{\upsilon}=\sum_{n=1}^{+\ns}\rbr{\sbr X_{\upsilon_{n}\wedge t}^{\sigma}-\sbr X_{\upsilon_{n-1}\wedge t}^{\sigma}-\sbr X_{\upsilon_{n}\wedge t}^{\upsilon}+\sbr X_{\upsilon_{n-1}\wedge t}^{\upsilon}}^{2}.
\]
Denoting
\[
n(t):=\max\cbr{n\in\N:\upsilon_{n}\le t},\quad t\in[0,+\ns),
\]
we further estimate 
\begin{align}
\sbr{\sbr X^{\sigma}-\sbr X^{\upsilon}}_{t}^{\upsilon}= & \sum_{n=1}^{n(t)}\rbr{\sbr X_{\upsilon_{n}}^{\sigma}-\sbr X_{\upsilon_{n-1}}^{\sigma}-\rbr{X_{\upsilon_{n}}-X_{\upsilon_{n-1}}}^{2}}^{2}\nonumber \\
 & +\rbr{\sbr X_{t}^{\sigma}-\sbr X_{\upsilon_{n(t)}}^{\sigma}-\rbr{X_{t}-X_{\upsilon_{n(t)}}}^{2}}^{2}.\label{eq:zero}
\end{align}
Next, denoting 
\[
m(n):=\max\cbr{m\in\N:\sigma_{m}\le\upsilon_{n}},\quad n\in\N,
\]
for $n\in\N\setminus\cbr 0$ we have 
\begin{align*}
\sbr X_{\upsilon_{n}}^{\sigma}-\sbr X_{\upsilon_{n-1}}^{\sigma} & =\rbr{X_{\upsilon_{n}}-X_{\sigma_{m\rbr{n-1}}}}^{2}-\rbr{X_{\upsilon_{n-1}}-X_{\sigma_{m\rbr{n-1}}}}^{2}\\
 & =\rbr{X_{\upsilon_{n}}-X_{\upsilon_{n-1}}}\rbr{X_{\upsilon_{n}}+X_{\upsilon_{n-1}}-2X_{\sigma_{m\rbr{n-1}}}}
\end{align*}
(this may be proven by considering two possible cases: $\upsilon_{n}=\sigma_{m(n)}>\upsilon_{n-1}\ge\sigma_{m(n-1)}$
and $\upsilon_{n}\ge\upsilon_{n-1}\ge\sigma_{m(n)}=\sigma_{m(n-1)}$)
so 
\begin{equation}
\sbr X_{\upsilon_{n}}^{\sigma}-\sbr X_{\upsilon_{n-1}}^{\sigma}-\rbr{X_{\upsilon_{n}}-X_{\upsilon_{n-1}}}^{2}=\rbr{X_{\upsilon_{n}}-X_{\upsilon_{n-1}}}\rbr{2X_{\upsilon_{n-1}}-2X_{\sigma_{m\rbr{n-1}}}}.\label{eq:jeden}
\end{equation}
Similarly, 
\begin{equation}
\sbr X_{t}^{\sigma}-\sbr X_{\upsilon_{n(t)}}^{\sigma}-\rbr{X_{t}-X_{\upsilon_{n(t)}}}^{2}=\rbr{X_{t}-X_{\upsilon_{n(t)}}}\rbr{2X_{\upsilon_{n(t)}}-2X_{\sigma_{m(n(t))}}}.\label{eq:dwa}
\end{equation}
Plugging in (\ref{eq:zero}) equalities (\ref{eq:jeden}) and (\ref{eq:dwa}),
and using the estimates 
\[\left|2X_{\upsilon_{n-1}}-2X_{\sigma_{m\rbr{n-1}}}\right|\le2\delta, \quad
\left|2X_{\upsilon_{n(t)}}-2X_{\sigma_{m(n(t))}}\right|\le2\delta, \]
which stem from \eqref{fin_cov_wie}, we get (\ref{eq:q_var_estim}).
\end{proof}

For a positive real number $d$ and $r\in[0,d)$
let us consider the grid $d\cdot\Z+r=\cbr{d\cdot n+r:n\in\Z}$. For
a real process $X$ let now $\tau(X,d,r)=\rbr{\tau_{n}(X,d,r)}$ be
a sequence of times $\tau_{n}=\tau_{n}(X,d,r)$ defined as:
$\tau_{0}\equiv0$ and for $n=1,2,\ldots$ 
\[
\tau_{n}=
\begin{cases} 
\inf\cbr{t>\tau_{n-1}:X_{t}\in\rbr{d\cdot\Z+r}\setminus\cbr{X_{\tau_{n-1}}}} \text{ if } \tau_{n-1}< +\ns; \\
+ \ns \text{ if } \tau_{n-1} = +\ns.
\end{cases}
\]
If $X$ is continuous w.i.e. then by Lemma \ref{stop_times}, $\tau_n$, $n \in \N$, is a stopping time. 

Moreover, if $X$ is a martingale then $\tau(X,d,r)$ is a fine cover of the process $X$ with accuracy $d>0$ w.i.e. (since the property of $\rbr{t, \omega} \in [0, +\ns)\times \Omega$ that the trajectory $[0, +\ns) \ni s \mapsto X_s(\omega)$ is real and continuous on $[0, t]$ is instantly enforceable). However, the sequence $\tau(X,d,r)$ may be not proper for all $\omega$. To avoid such a situation we modify $\tau_n$ by setting: $\tau_n(\omega) = +\ns$ if $X$ is not real and continuous on $[0, \tau_n(\omega))$. Such modified $\tau_n$ is also a stopping time (the proof is almost the same as the proof of  Lemma \ref{stop_times}) and a sequence of such modified times is proper and is a fine cover of the process $X$ with accuracy $d>0$ w.i.e. We will also denote it by $\tau(X,d,r)$ and call the \emph{Lebesgue sequence of stopping times} for $X$ and the grid $d\cdot\Z+r$.

To state the next proposition we need two more definitions.  
\begin{definition} Let $\sigma$ be a stopping time. By the \emph{locally uniform convergence
of the sequence of processes $\rbr{Y^{m}}$ on the random interval
$[0,\sigma]\setminus\cbr{+\ns}$, with instant enforcement (w.i.e.)}, to
the process $Y$, we mean the fact that the property of $(t,\omega)\in[0,+\ns)\times\Omega$
that $\sup_{s\in[0,\sigma\wedge t]}\left|Y_{s}^{m}(\omega)-Y_{s}(\omega)\right|\rightarrow0$ holds w.i.e.

By the \emph{locally uniform convergence of the sequence of processes $\rbr{Y^{m}}$
w.i.e.} to the process $Y$ we mean the fact that  the property of $(t,\omega)\in[0,+\ns)\times\Omega$
that $\sup_{s \in [0, t]}\left|Y_{s}^{m}(\omega)-Y_{s}(\omega)\right|\rightarrow0$ holds w.i.e. 
\end{definition}
Now we are ready to state and prove a proposition on the existence of martingale quadratic variation. 

\begin{proposition} \label{103} Let $X$ be a martingale. There exists a real continuous process $\sbr X$ such that if  $\rbr{\delta_m}$ is a sequence of positive reals such that $\sum_{m=0}^{+\ns}\delta_{m}<+\ns$ and 
$\rbr{\sigma^{m}}$ is a sequence of sequences $\sigma^{m}=\rbr{\sigma_{n}^{m}}_n$
of stopping times, such that $\sigma^{m}$ is proper and is a fine cover of $X$
with accuracy $\delta_{m}$ w.i.e. then, for any $M\in(0,+\ns)$ and $\sigma(M)=\sigma(X,M)$ defined by (\ref{eq:sigmaM}),
the processes $\sbr X^{\sigma^{m}}$ converge locally
uniformly on the random interval $[0,\sigma(M)]\setminus\cbr{+\ns}$
with instant enforcement to the process $\sbr X$. As a result, the processes $\sbr X^{\sigma^{m}}$ converge locally
uniformly w.i.e. to the process $\sbr X$.
\end{proposition}

\begin{proof} 
First we consider $\tau^m = \tau\rbr{X,2^{-m},0}$, $m \in \N$, -- the Lebesgue sequences of stopping times for $X$ and the grid $2^{-m}\Z$. Let us fix $M\in(0,+\ns)$ and let $E$ be the instantly enforceable set of $(t,\omega) \in [0, +\ns)\times \Omega$ where the trajectory $[0, t] \ni s \mapsto X_s(\omega)$ is continuous. 

Since $\tau^{m+1}$ is the same as the non-decreasing rearrangement of $\tau^{m}$ and
$\tau^{m+1}$ (all stopping times from the sequence $\tau^{m}$ also
appear in the sequence $\tau^{m+1}$) and since $\tau^{m}$ is a fine
cover of $X$ with accuracy $2^{-m}$ on the set $E$, by Lemma \ref{102} we have
\begin{equation} \label{eq:q_var_estim-1}
\sbr{\sbr X^{\tau^{m}}(\omega)-\sbr X^{\tau^{m+1}}(\omega)}_{t}^{\tau^{m+1}}\le4\cdot2^{-2m}\sbr X_{t}^{\tau^{m+1}}(\omega) \text{ for } (t, \omega) \in E.
\end{equation}
By Fact \ref{100} the difference 
\begin{align*}
 Y_{t}^{m}   := &\sbr X_{t\wedge\sigma(M)}^{\tau^{m+1}}-\sbr X_{t\wedge\sigma(M)}^{\tau^{m}}\\
  = &\rbr{X_{t\wedge\sigma(M)}-X_{0}}^{2}-\sbr X_{t\wedge\sigma(M)}^{\tau^{m}}-\rbr{\rbr{X_{t\wedge\sigma(M)}-X_{0}}^{2}-\sbr X_{t\wedge\sigma(M)}^{\tau^{m+1}}}.
\end{align*}
is a difference of two martingales stopped at $\sigma(M)$, thus a
martingale. Recall a definition of the supremum process (of a generalized
process) and consider $\rbr{Y^{m}}^{*}$. Now, since $\tau^{m+1}$,
$m\in\N$, is a fine cover of $X$ with accuracy $2^{-m-1}$ on $E$, we have
\[
\rbr{Y^{m}}_{t}^{*}\le\rbr{Y^{m}}_{t}^{\tau^{m+1},*}+2\cdot 2^{-2m-2}
\]
on this set, and by this and Fact \ref{99} (discrete BDG inequality) we have
\begin{align*}
\overline{\E}\rbr{Y^{m}}^{*} & \le\overline{\E}\rbr{Y^{m}}^{\tau^{m+1},*}+2^{-2m-1} \le6\overline{\E}\sqrt{\sbr{Y^{m}}^{\tau^{m+1}}}+2^{-2m-1}.
\end{align*}
Further, using (\ref{eq:q_var_estim-1}), the elementary estimate
$\sqrt{x}\le\frac{1}{2}+\frac{1}{2}x$ ($x\ge0$) and the It\^o isometry
(Fact \ref{101}) we have 
\begin{align}
\overline{\E}\rbr{Y^{m}}^{*} & \le6\overline{\E}\sqrt{4\cdot2^{-2m}\sbr X^{\tau^{m+1}}_{\cdot \wedge\sigma(M)}}+2^{-2m-1}\nonumber \\
 & \le6\cdot2^{-m}\rbr{1+\overline{\E}\sbr X^{\tau^{m+1}}_{\cdot \wedge\sigma(M)}}+2^{-2m-1}\nonumber \\
 & \le6\cdot2^{-m}\rbr{1+\overline{\E}\rbr{X-X_{0}}_{\cdot\wedge\sigma(M)}^{2}}+2^{-2m-1}\nonumber \\
 & \le6\cdot2^{-m}\rbr{1+4M^{2}}+2^{-2m-1}\nonumber \\
 & \le7\rbr{1+4M^{2}}2^{-m}.\label{eq:out_exp_estim}
\end{align}
Now let $B\subseteq[0,+\ns)\times\Omega$ be the set of pairs $(t, \omega)$ where the sequence
of processes $\sbr X^{\tau^{m}}$, $m\in\N$, does not converge uniformly
on $\sbr{0,\sigma(M)\wedge t}\setminus\cbr{+\ns}$. Let us fix $\varepsilon>0$. For each $(t,\omega)\in B$ we
have 
\[
\varepsilon\sum_{m=0}^{+\ns}\rbr{Y^{m}}_{t}^{*}(\omega)=+\ns\ge{\bf 1}_{B}(t,\omega).
\]
By (\ref{eq:out_exp_estim}) there exists a nonnegative supermartingale
$Z^{m}$ such that $Z_{0}^{m}\le8\rbr{1+4M^{2}}2^{-m}$ and $Z_{t}^{m}(\omega)\ge\rbr{Y^{m}}_{t}^{*}(\omega)$
for each $(t,\omega)\in[0,+\ns)\times\Omega$. Hence 
\[
U^{\varepsilon}:=\varepsilon\cdot\sum_{m=0}^{+\ns}Z^{m}
\]
is a nonnegative supermartingale such that $U_{0}^{\varepsilon}\le8\rbr{1+4M^{2}}\varepsilon\sum_{m=0}^{+\ns}2^{-m}=16\rbr{1+4M^{2}}\varepsilon$
and for each $(t,\omega)\in B$
\[
U_{t}^{\varepsilon}(\omega)=+\ns>{\bf 1}_{B}(t,\omega).
\]
Since $\varepsilon$ may be as close to $0$ as we wish, we get that
the set $B$ is instantly blockable. 

Let $[X]$ denote any real continuous process to which $\sbr X^{\tau^{m}}$
converges locally uniformly on $[0,\sigma(M)]\setminus\cbr{+\ns}$ w.i.e. for all 
$M=1,2,\ldots$ (we may take for example
$[X]_{t}(\omega):=\lim_{m\ra+\ns}\sbr X_{t}^{\tau^{m}}(\omega)$ if
the limit exists and $[X]_{t}(\omega):=0$ if the limit does not exist). 

For $m\in\N$ let now $\upsilon^{m}$ be the non-decreasing rearrangement
of the stopping times from both sequences $\sigma^{m}$ and $\tau^{m}$
with redundancies deleted. Reasoning similarly as for $\tau^{m}$
and $\tau^{m+1}$ we infer that for the differences 

\[
R^{m}:=\sbr X^{\sigma^{m}}_{\cdot \wedge\sigma(M)}-\sbr X^{\upsilon^{m}}_{\cdot \wedge\sigma(M)},\quad V^{m}:=\sbr X^{\tau^{m}}_{\cdot \wedge\sigma(M)}-\sbr X^{\upsilon^{m}}_{\cdot \wedge\sigma(M)}
\]
one has 
\begin{equation}
\overline{\E}\rbr{R^{m}}^{*}\le6\delta_{m}\rbr{1+4M^{2}}+2\delta_{m}^{2}\label{eq:ineqlan}
\end{equation}
and 
\begin{equation}
\overline{\E}\rbr{V^{m}}^{*}\le6\cdot2^{-m}\rbr{1+4M^{2}}+2\cdot2^{-2m}.\label{eq:ineqtwo}
\end{equation}
Now, if $D(M)\subseteq[0,+\ns)\times\Omega$ is the set of pairs $(t, \omega)$  where the sequence
of processes $\sbr X^{\sigma^{m}}$, $m\in\N$, does not converge uniformly
to $[X]$ on $\sbr{0,\sigma(M)\wedge t}\setminus\cbr{+\ns}$ then for each
$\varepsilon>0$ and $(t,\omega)\in D(M)$
\begin{equation}
\varepsilon\sum_{m=0}^{+\ns}\rbr{R^{m}}_{t}^{*}(\omega)+\varepsilon\sum_{m=0}^{+\ns}\rbr{V^{m}}_{t}^{*}(\omega)=+\ns\ge{\bf 1}_{D(M)}(t,\omega).\label{eq:ineqthree}
\end{equation}
Inequalities (\ref{eq:ineqlan}), (\ref{eq:ineqtwo}) and (\ref{eq:ineqthree})
imply the existence of a nonnegative supermartingale which starts from
the initial capital smaller than $$6\varepsilon(1+4M^{2})\sum_{m=0}^{+\ns}\rbr{\delta_{m}+2^{-m}}+2\varepsilon\sum_{m=0}^{+\ns}\rbr{\delta_{m}^{2}+2^{-2m}}$$
and attains value $+\ns$ on the set $D(M)$. Thus, since $\varepsilon$
may be arbitrary close to $0$, $D(M)$ is instantly blockable. 

To obtain locally uniform convergence w.i.e. of $\sbr X^{\sigma^{m}}$ to $\sbr{X}$ we consider $D:= \bigcup_{M=1}^{+\ns} D(M)$. $D$ is instantly blockable and on its complement we have the desired convergence.
\end{proof}
\begin{definition} \label{q_var_def}
By \emph{quadratic variation} of a martingale $X$ we will mean any process 
which satisfies the thesis of Proposition \ref{103}. $[X]$ is real, continuous and increasing w.i.e. Any two processes satisfying the thesis of Proposition \ref{103} are equal w.i.e.
\end{definition}

A direct consequence of Lemma \ref{100}, Proposition \ref{103} and Definition \ref{q_var_def} is the following fact.

\begin{fact} \label{q_var_mart}
If $X$ is a martingale then the process $\rbr{X-X_0}^2-[X]$ is also a martingale.
\end{fact}

We also have the following fact.

\begin{fact} \label{q_var_mart0}
If $X$ is a martingale, $X_0 =0$ and $[X]=0$ w.i.e. then $X =0$ w.i.e.
\end{fact}
\begin{proof}
By Fact \ref{q_var_mart}, $X^2$ is a martingale starting from $0$. It is nonnegative, thus Corollary \ref{ebar_est} yields that $\overline{\E}X^2 = 0$. For any $n \in \N$ there exists a nonnegative supermartingale $U^n$ such that $U^n_0 \le 1/n^2$ and $U^n \ge X^2$. Thus, $U = \liminf_{n \ra +\ns} n \cdot U^n$ is a nonnegative supermartingale such that $U_0 = 0$ and for any $(t, \omega) \in [0,+\ns)\times\Omega$ such that $X_t(\omega) \neq 0$, $U_t(\omega)=+\ns$, hence $X = 0$ w.i.e.
\end{proof}

\subsubsection{Quadratic covariation} \label{q_cov}

Now, let $X$ and $Y$ be two martingales. Naturally, $X+Y$ and $X-Y$ are also martingales. Let $\rbr{\tau(X,2^{-m},0)}$ and $\rbr{\tau(Y,2^{-m},0)}$ be two sequences of the Lebesgue sequences of stopping times for $X$
and $Y$ respectively (and the grids $2^{-m}\cdot\Z$). Let $\upsilon^{m}$ be the non-decreasing rearrangement
of the stopping times from both sequences $\tau(X,2^{-m},0)$ and $\tau(Y,2^{-m},0)$,
with redundancies deleted. $\upsilon^{m}$ is also a proper sequence of stopping times and finely covers {w.i.e.} both -- $X+Y$ and $X-Y$ -- with accuracy $2^{-m+1}$. By Proposition \ref{103} we get that for any real $M>0$, $\sbr{X+Y}^{\upsilon^{m}}$  and $\sbr{X-Y}^{\upsilon^{m}}$  converge locally uniformly on $\sbr{0, \sigma(X+Y, M) \wedge \sigma(X-Y, M)}\setminus\cbr{+\ns}$ w.i.e. to the quadratic variations $\sbr{X+Y}$ and $\sbr{X-Y}$ respectively. The difference 
\[
[X,Y] := \frac{1}{4}[X+Y]-\frac{1}{4}[X-Y]
\]
is called \emph{the quadratic covariation of $X$ and $Y$}. Substituting $a = X_{\upsilon^{m}_{n}\wedge t}-X_{\upsilon^{m}_{n-1}\wedge t}$, $b = Y_{\upsilon^{m}_{n}\wedge t}-Y_{\upsilon^{m}_{n-1}\wedge t}$ in the identity $\frac{1}{4}(a+b)^2 - \frac{1}{4}(a-b)^2=a\cdot b$ we get that $[X,Y]$ is the limit of the \emph{simple quadratic covariation processes along $\rbr{\upsilon^{m}}$}:
\begin{equation} \label{cov0}
[X,Y]^{\upsilon^{m}}_t := \sum_{n=1}^{+\ns}\rbr{X_{\upsilon^{m}_{n}\wedge t}-X_{\upsilon^{m}_{n-1}\wedge t}}\rbr{Y_{\upsilon^{m}_{n}\wedge t}-Y_{\upsilon^{m}_{n-1}\wedge t}},\quad t\in[0,+\ns), 
\end{equation}
which converge to $[X,Y]$ locally uniformly on $\sbr{0, \sigma(X+Y, M) \wedge \sigma(X-Y, M)}\setminus\cbr{+\ns}$  w.i.e. for any  real $M>0$.

\subsection{It\^o's isometry}

Using Fact \ref{101} (It\^o's isometry for simple quadratic variation) and the just
proven Proposition \ref{103} we can obtain It\^o's isometry  for quadratic variation.

\begin{fact}[It\^o's isometry  for quadratic variation] \label{1011} Let $X$ be a martingale and $\sbr{X}$ its quadratic variation. We have 
\[
\overline{\E}\rbr{X-X_{0}}^{2}=\overline{\E}\sbr X.
\]
\end{fact}
\begin{proof}
Let $\rbr{\tau^{m}}=\rbr{\tau(X,2^{-m},0)}$ be the
sequence of the Lebesgue sequences of stopping times for $X$ and
the grids $2^{-m}\cdot\Z$.
For any $m=0,1,2,\ldots$, by the It\^o isometry for simple quadratic variation (Fact \ref{101})
we have 
$
\overline{\E}{\rbr{X-X_{0}}^{2}} = \overline{\E} {\sbr X^{\tau^{m}} }
$
which yields
\begin{equation}
\overline{\E}{\rbr{X-X_{0}}^{2}}  = \liminf_{m\ra+\ns}\overline{\E} {\sbr X^{\tau^{m}} }.\label{eq:ito_cont_1}
\end{equation}
Since $\liminf_{m\ra+\ns}\sbr X^{\tau^{m}}=[X]$ w.i.e. 
we have $\overline{\E}{[X]}=\overline{\E}\liminf_{m\ra+\ns}{\sbr X^{\tau^{m}} }$
and by (\ref{eq:ito_cont_1}) and the Fatou lemma (Lemma \ref{98}) we get
\[
\overline{\E}{\rbr{X-X_{0}}^2}\ge \overline{\E}\liminf_{m\ra+\ns}{\sbr X^{\tau^{m}} } = \overline{\E}{[X]}.
\]

To prove the upper bound let us fix $\varepsilon,M\in (0,+\ns)$. Let $\sigma(M)=\sigma(X,M)$
be defined by (\ref{eq:sigmaM}) and 
\[
\rho^{m}(\varepsilon):=\inf\cbr{t\ge0:\left|{\sbr X_{t}}-{\sbr X_{t}^{\tau^{m}}}\right|\ge\varepsilon},
\]
where we take $[X]=\liminf_{m\ra+\ns}\sbr X^{\tau^{m}}$. 
By Lemma \ref{stop_times}, $\rho^{m}(\varepsilon)$ is a stopping time ($[X]$ is continuous w.i.e.).

Similarly as in the proof of the It\^o isometry for simple quadratic variation we define a trading
strategy $G^{m,\varepsilon,M}=\rbr{\varepsilon,\tau^{m},\rbr{g_{n}^{m,\varepsilon,M}}}$
in the following way: 
\begin{align*}
g_{n}^{m,\varepsilon,M}(\omega): & =\begin{cases}
2\rbr{X_{\tau_{n}}(\omega)-X_{0}(\omega)}  & \text{if }n\in\N\text{ and }\tau_{n}^{m}<\sigma(M)\wedge\rho^{m}(\varepsilon),\\
0 & \text{if }n\in\N\text{ and }\tau_{n}^{m}\ge\sigma(M)\wedge\rho^{m}(\varepsilon).
\end{cases}
\end{align*}
Functions $g_{n}^{m,\varepsilon,M}$ are globally
bounded (by $2M$) and ${\cal F}_{\tau_{n}^{m}}$-measurable.
The pathwise limit 
\[ G_{t}^{m,\varepsilon,X}(\omega):=\lim_{M\ra+\ns}\rbr{G^{m,\varepsilon,M}\cdot X}_{t}(\omega), \quad
(t,\omega)\in[0,+\ns)\times\Omega,
\] 
is a martingale since $\rbr{G^{m,\varepsilon,M}\cdot X}_{s}(\omega)$, $s \in [0, t]$, 
is the same for all $M>\sup_{s\in[0,t]}\left|X_{s}(\omega)\right|$, ($\sup_{s\in[0,t]}\left|X_{s}(\omega)\right|$ is finite w.i.e.).
Moreover, if $Y$ is a nonnegative supermartingale dominating ${\sbr X}$
and such that $Y_{0}\le\varepsilon+\overline{\E}{\sbr X}$
then by the definition of the stopping time $\rho^{m}(\varepsilon)$,
$\varepsilon+Y_{\cdot\wedge\rho^{m}(\varepsilon)}$ is a nonnegative
supermartingale dominating ${\sbr X_{\cdot\wedge\rho^{m}(\varepsilon)}^{\tau^{m}}}$.
Next, ${\varepsilon+Y_{\cdot\wedge\rho^{m}(\varepsilon)}+G^{m,\varepsilon,X}}$ is a process dominating ${\rbr{X-X_{0}}_{\cdot\wedge\rho^{m}(\varepsilon)}^{2}}$ and by Corollary \ref{ebar_est} 
\[
\overline{\E} \cbr{\varepsilon+Y_{\cdot\wedge\rho^{m}(\varepsilon)}+G^{m,\varepsilon,X}} = \varepsilon+Y_{0}.
\] 
Proceeding to the lower limit we have that 
$$\liminf_{m\ra+\ns}\cbr{{\varepsilon+Y_{\cdot\wedge\rho^{m}(\varepsilon)}}+2G^{m,\varepsilon,X}}
$$
is a nonnegative process which dominates
\[
\liminf_{m\ra+\ns}{\rbr{X-X_{0}}_{\cdot\wedge\rho^{m}(\varepsilon)}^{2}}={\rbr{X-X_{0}}^{2}} \quad w.i.e. 
\]
(since $\sbr X^{\tau^{m}}$ converges locally uniformly w.i.e.). Hence, by the Fatou lemma,
\[
\overline{\E}{\rbr{X-X_{0}}^{2}}\le \liminf_{m \ra +\ns}\overline{\E} \cbr{\varepsilon+Y_{\cdot\wedge\rho^{m}(\varepsilon)}+G^{m,\varepsilon,X}} = {\varepsilon+Y_{0}}\le {2\varepsilon+\overline{\E}\sbr X}
\]
which gives the desired bound by letting $\varepsilon\ra0+$.

\end{proof}

\subsection{BDG inequalities}

Using Fact \ref{99} (BDG inequalities for simple quadratic variation) and Proposition \ref{103} we obtain the following proposition.

\begin{proposition}[BDG inequalities  for quadratic variation] \label{104} Let $X$ be
a martingale and $[X]$ be its quadratic variation. For any $p\ge1$
there exist finite, positive constants $c_{p}$ and $C_{p}$ such
that 
\[
c_{p}\overline{\E}\sbr X^{p/2}\le\overline{\E}\rbr{\rbr{X-X_{0}}^{*}}^{p}\le C_{p}\overline{\E}\sbr X^{p/2}.
\]
In the case $p>1$ one may take $C_{p}=6^{p}(p-1)^{p-1}$ and $c_{p}=1/C_{p}$,
while in the case $p=1$ one may take $C_{p}=6$ and $c_{p}=1/3$.
\end{proposition}
\begin{proof} The proof is similar to the proof of It\^o's isometry for quadratic variation. Let $\rbr{\tau^{m}}=\rbr{\tau(X,2^{-m},0)}$ be the
sequence of the Lebesgue sequences of stopping times for $X$ and
the grids $2^{-m}\cdot\Z$.
For any $m=0,1,2,\ldots$, by the BDG inequality for simple quadratic variation (Fact \ref{99}),
we have 
\[
\overline{\E}\rbr{\rbr{X-X_{0}}^{*}}^{p}\ge\overline{\E}\rbr{\rbr{X-X_{0}}^{\tau^{m},*}}^{p}\ge c_{p}\overline{\E} \rbr{\sbr X^{\tau^{m}} }^{p/2}
\]
which yields
\begin{equation}
\overline{\E}\rbr{\rbr{X-X_{0}}^{*}}^{p}\ge c_{p}\liminf_{m\ra+\ns}\overline{\E}\rbr{\sbr X^{\tau^{m}} }^{p/2}.\label{eq:bdg_cont_1}
\end{equation}
Since $\liminf_{m\ra+\ns}\sbr X^{\tau^{m}}=[X]$ w.i.e. (understood
for all $(t,\omega)\in[0,+\ns)\times\Omega$ as the property that
$\liminf_{m\ra+\ns}\sbr X_{t}^{\tau^{m}}(\omega)=[X]_{t}(\omega)$)
we have $\overline{\E}\rbr{[X]}^{p/2}=\overline{\E}\liminf_{m\ra+\ns}\rbr{\sbr X^{\tau^{m}} }^{p/2}$
and by (\ref{eq:bdg_cont_1}) and the Fatou lemma (Lemma \ref{98}) we get
\[
\overline{\E}\rbr{\rbr{X-X_{0}}^{*}}^{p}\ge c_{p}\overline{\E}\liminf_{m\ra+\ns} \rbr{\sbr X^{\tau^{m}} }^{p/2} =c_{p}\overline{\E}\rbr{[X]}^{p/2}.
\]

To prove the upper bound let us fix $\varepsilon,M\in(0,+\ns)$. Let $\sigma(M)=\sigma(X,M)$
be defined by (\ref{eq:sigmaM}) and 
\[
\rho^{m}(\varepsilon):=\inf\cbr{t\ge0:\left|\rbr{\sbr X_{t}}^{p/2}-\rbr{\sbr X_{t}^{\tau^{m}}}^{p/2}\right|\ge\varepsilon},
\]
where we take $[X]=\liminf_{m\ra+\ns}\sbr X^{\tau^{m}}$. By Lemma \ref{stop_times}, $\rho^{m}(\varepsilon)$ is a stopping time ($[X]$ is continuous w.i.e.).
 Similarly
as in the proof of the BDG inequality for simple quadratic variation we define a trading
strategy $G^{m,\varepsilon,M}=\rbr{\varepsilon,\tau^{m},\rbr{g_{n}^{m,\varepsilon,M}}}$
in the following way: $x_{n}=X_{\tau_{n}^{m}}(\omega)-X_{0}$ and
\begin{align*}
g_{n}^{m,\varepsilon,M}(\omega): & =\begin{cases}
g_{n} & \text{if }n\in\N\text{ and }\tau_{n}^{m}<\sigma(M)\wedge\rho^{m}(\varepsilon),\\
0 & \text{if }n\in\N\text{ and }\tau_{n}^{m}\ge\sigma(M)\wedge\rho^{m}(\varepsilon),
\end{cases}
\end{align*}
where $g_{n}$ for the given sequence $\rbr{x_{n}}$ is defined as
in (\ref{eq:fgdef}). Functions $g_{n}^{m,\varepsilon,M}$ are globally
bounded (by constants depending on $m,n$ and $M$) and ${\cal F}_{\tau_{n}^{m}}$-measurable.
The pathwise limit 
\[ G_{t}^{m,\varepsilon,X}(\omega):=\lim_{M\ra+\ns}\rbr{G^{m,\varepsilon,M}\cdot X}_{t}(\omega), \quad
(t,\omega)\in[0,+\ns)\times\Omega,
\] 
is a martingale since $\rbr{G^{m,\varepsilon,M}\cdot X}_{t}(\omega)$
is the same for all $M>\sup_{s\in[0,t]}\left|X_{s}(\omega)\right|$, ($\sup_{s\in[0,t]}\left|X_{s}(\omega)\right|$ is finite w.i.e.).
Moreover, if $Y$ is a nonnegative supermartingale dominating $\rbr{\sbr X}^{p/2}$
and such that $Y_{0}\le\varepsilon+\overline{\E}\rbr{\sbr X}^{p/2}$
then by the definition of the stopping time $\rho^{m}(\varepsilon)$,
$\varepsilon+Y_{\cdot\wedge\rho^{m}(\varepsilon)}$ is a nonnegative
supermartingale dominating $\rbr{\sbr X_{\cdot\wedge\rho^{m}(\varepsilon)}^{\tau^{m}}}^{p/2}$.
Next, by (\ref{eq:BDG>1}), $C_{p}\rbr{\varepsilon+Y_{\cdot\wedge\rho^{m}(\varepsilon)}}+2G^{m,\varepsilon,X}$
is a process dominating $\rbr{\rbr{X-X_{0}}_{\cdot\wedge\rho^{m}(\varepsilon)}^{\tau^{m},*}}^{p}$
(this follows from the application of (\ref{eq:BDG>1}) to the sequence
$\tilde{x}_{k}=X_{\tau_{k}^{m}\wedge t}-X_{0}$, $k\in\N$) and by Corollary \ref{ebar_est} 
\[
\overline{\E} \cbr{C_{p}\rbr{\varepsilon+Y_{\cdot\wedge\rho^{m}(\varepsilon)}}+2G^{m,\varepsilon,X}} =  C_{p}\rbr{\varepsilon+Y_{0}}.
\]
Proceeding to the lower limit we have that 
$$\liminf_{m\ra+\ns}\cbr{C_{p}\rbr{\varepsilon+Y_{\cdot\wedge\rho^{m}(\varepsilon)}}+2G^{m,\varepsilon,X}}$$ is a nonnegative supermartingale dominating 
\[
\liminf_{m\ra+\ns}\rbr{\rbr{X-X_{0}}_{\cdot\wedge\rho^{m}(\varepsilon)}^{\tau^{m},*}}^{p}=\rbr{\rbr{X-X_{0}}^{*}}^{p} \quad w.i.e. 
\]
(let us notice that since $\tau^{m}$ finely covers $X$ with accuracy
$2^{-m}$ w.i.e., we have 
\[
\rbr{X-X_{0}}^{\tau^{m},*}\le\rbr{X-X_{0}}^{*}\le\rbr{X-X_{0}}^{\tau^{m},*}+2^{-m} \quad w.i.e.
\]
and since $\sbr X^{\tau^{m}}$ converges locally uniformly w.i.e., 
$\rho^{m}(\varepsilon)(\omega) \ge t$ as $m\ra+\ns$ w.i.e.) hence, by the Fatou lemma,
\begin{align*}
\overline{\E}\rbr{\rbr{X-X_{0}}^{*}}^{p}&\le \liminf_{m \ra +\ns} \overline{\E} \cbr{C_{p}\rbr{\varepsilon+Y_{\cdot\wedge\rho^{m}(\varepsilon)}}+2G^{m,\varepsilon,X}} \\ & =C_{p}\rbr{\varepsilon+Y_{0}}\le C_{p}\rbr{2\varepsilon+\overline{\E}\rbr{\sbr X}^{p/2}},
\end{align*}
which gives the desired bound by letting $\varepsilon\ra0+$.
\end{proof}

\section{Model-free stochastic integral -- definition and its quadratic variation}
\subsection{Model-free stochastic integral -- definition}

If $G=\rbr{c,\rbr{\tau_{n}},\rbr{g_{n}}}$ is a simple trading strategy and $X$ is a martingale then $G \cdot X$ is again a martingale.  An almost immediate consequence of Proposition \ref{103} is the following fact.
\begin{fact} \label{q_var_simp_int}
Let $G=\rbr{c,\rbr{\tau_{n}},\rbr{g_{n}}}$, $H=\rbr{d,\rbr{\sigma_{n}},\rbr{h_{n}}}$ be simple trading strategies and $X$, $Y$ be martingales. Then the quadratic covariation of the martingales $G \cdot X$ and $H \cdot Y$ equals 
\[
\sbr{G \cdot X, H \cdot Y}_t = \int_0^t \rbr{G_s \cdot H_s}\cdot \dd [X, Y]_s \text{ w.i.e.,}
\]
where $G_t := \sum_{n=1}^{+\ns} g_{n-1} {\bf 1}_{\left[\tau_{n-1}, \tau_n \right)} (t)$, $H_t := \sum_{n=1}^{+\ns} h_{n-1} {\bf 1}_{\left[\sigma_{n-1}, \sigma_n \right)} (t)$ and the integral $\int_0^t G_s \cdot H_s\cdot \dd [X, Y]_s$ is understood as the (pathwise) Lebesque-Stieltjes integral.
\end{fact}
\begin{proof}[Sketch of a proof]
Let $\tilde{G} = \rbr{c,\rbr{\tilde{\tau}_{n}},\rbr{\tilde{g}_{n}}}$ be a modification of the trading strategy $G$ obtained in the following way. If for some $\omega \in \Omega$ and $n\in \N$, $\tau_n(\omega) = \tau_{n+1}(\omega) = \ldots$ then we set 
\[
\tilde{\tau}_{n} = \tilde{\tau}_{n+1} = \ldots = +\ns, \quad \tilde{g}_{n} = \tilde{g}_{n+1} = \ldots  = 0,
\]
otherwise nor $\tau_n$ neither $g_n$ are changed. This way we get a trading strategy satisfying for \emph{all} $(t,\omega) \in [0,+\ns) \times \Omega$, $\rbr{\tilde{G} \cdot X}_t(\omega)=\rbr{G \cdot X}_t(\omega)$ and such that for \emph{all} $\omega \in \Omega$, $\tau_n (\omega) \ra +\ns$.

Let $\tilde{H} = \rbr{d,\rbr{\tilde{\sigma}_{n}},\rbr{\tilde{h}_{n}}}$ be a similar modification of the strategy $H$.

Let us consider $m\in \N$ and  let $\tau(X,2^{-m},0)$, $\tau(Y,2^{-m},0)$, $\tau(\tilde{G} \cdot X,2^{-m},0)$ and $\tau(\tilde{H} \cdot Y,2^{-m},0)$ be the Lebesgue sequences of stopping times for $X$, 
$Y$, $\tilde{G} \cdot X$ and $\tilde{H} \cdot Y$ respectively (and the grid $2^{-m}\cdot\Z$), and $\upsilon^{m}$ be the non-decreasing rearrangement
of the stopping times from these sequences  and $\rbr{\tilde{\tau}_n}_n$, $\rbr{\tilde{\sigma}_{n}}_n$, with redundancies deleted. Since $\upsilon^{m}$ is a proper sequence of stopping times and finely covers both -- $X+Y$ and $X-Y$ w.i.e. with accuracy $2^{-m+1}$ -- we easily see that by \eqref{cov0} 
\begin{align}
& \sbr{G \cdot X, H \cdot Y}^{\upsilon^{m}}_t 
 = \sbr{\tilde{G} \cdot X, \tilde{H} \cdot Y}^{\upsilon^{m}}_t  \nonumber \\
& = \sum_{l=1}^{+\ns} G_{\upsilon^{m}_{l-1}  } \cdot H_{\upsilon^{m}_{l-1}  } \rbr{[X]_{\upsilon^{m}_l \wedge t}^{\upsilon^{m}} - [X]_{\upsilon^{m}_{l-1} \wedge t}^{\upsilon^{m}}}\rbr{[Y]_{\upsilon^{m}_l \wedge t}^{\upsilon^{m}} - [Y]_{\upsilon^{m}_{l-1} \wedge t}^{\upsilon^{m}}} \label{cov}
\end{align}
and by Proposition \ref{103}, $\sbr{G \cdot X, H \cdot Y}^{\upsilon^{m}}$ tends locally uniformly w.i.e. Moreover the limit is equal to the process 
$$\int_0^t \rbr{G_s \cdot H_s} \cdot \dd [X,Y]_s$$
since the (random) functions $t \mapsto G_t$, $t \mapsto H_t$ are constant on intervals of the form $\left[\tilde{\tau}_{n-1}, \tilde{\tau}_{n} \right)$, $\left[\tilde{\sigma}_{n-1}, \tilde{\sigma}_{n} \right)$ respectively (we apply the convention that $[+\ns, +\ns) = \emptyset$).
\end{proof}

In particular, taking in Fact \ref{q_var_simp_int} $G = H$, we get that 
\[
\sbr{G \cdot X}_t = \sum_{n=1}^{+\ns} g_{n-1}^2 \rbr{[X]_{\sigma_n \wedge t} - [X]_{\sigma_{n-1} \wedge t}} = \int_0^t G_s^2 \cdot \dd [X]_s \text{  w.i.e.}
\]

Now, having at hand Fact \ref{q_var_simp_int}, Remark \ref{conservatism} and BDG inequalities we are going to extend the definition of the integral with the martingale integrator $X$. 

Similarly as in \cite{LochPerkPro:2018} we equip the family of simple trading strategies with the following pseudo-distance. For two simple trading strategies $G=\rbr{c,\rbr{\tau_{n}},\rbr{g_{n}}}$ and $H=\rbr{d,\rbr{\sigma_{n}},\rbr{h_{n}}}$ we define
\[
d_{QV,X,loc} \rbr{G, H} := \sum_{N=1}^{\ns} 2^{-N} \overline{\E} \rbr{ \int_0^{\sigma(X, N)} \rbr{G_s - H_s}^2 \dd \sbr{X}_s}^{1/2}, 
\]
where $G_t := \sum_{n=1}^{+\ns} g_{n-1} {\bf 1}_{\left[\tau_{n-1}, \tau_n \right)} (t)$, $H_t := \sum_{n=1}^{+\ns} h_{n-1} {\bf 1}_{\left[\sigma_{n-1}, \sigma_n \right)} (t)$, $\sigma(X,N)$
is defined by (\ref{eq:sigmaM})  and the integral $ \int_0^{\sigma(X, N)} \rbr{G_s - H_s}^2 \dd \sbr{X}_s$ is understood as the (pathwise) Lebesque-Stieltjes integral. 

Similarly one can also define $d_{QV,X,loc} \rbr{G, H}$ for any two generalized processes $G$ and $H$ such that for any $t \ge0$ and $\omega \in \Omega$, the trajectories $s \mapsto G_s(\omega)$ and $s \mapsto H_s(\omega)$ are Borel measurable and real on the whole interval $[0, t]$ w.i.e. (as a property of $(t, \omega)$). Let us define the space of such generalized processes more formally.
\begin{definition}
By $\mathcal R$ we denote  the space of generalized processes  $G$ such that for any $t \ge0$ and $\omega \in \Omega$, the trajectories $s \mapsto G_s(\omega)$ are Borel measurable and real on the whole interval $[0, t]$ w.i.e. (as a property of $(t, \omega)$).
\end{definition}
We restrict our considerations to the space $\mathcal R$ since in order to calculate $d_{QV,X,loc}  \rbr{G, H}$ we need to calculate integrals $\int_0^{\sigma(X, N)} \rbr{G_s - H_s}^2 \dd \sbr{X}_s$  which remain undefined when $G_s = +\ns$ and $H_s = + \ns$ or  $G_s = -\ns$ and $H_s = - \ns$ for some $s \in [0, \sigma(X, N)]$, or the difference $G_s - H_s$ is not Borel measurable. However, when it occurs on an instantly blockable set, this does not affect the value of  $d_{QV,X,loc}  \rbr{G, H}$. From the Minkowski inequality for the Stieltjes integrals it follows that  $d_{QV,X,loc}$ satisfies the triangle inequality. We call $d_{QV,X,loc}$ a pseudo-distance since, for example, the paths $s \mapsto G_s$ and $s \mapsto H_s$, $s\in [0, +\ns)$, may differ on the intervals where the martingale $X$ is constant, but still $d_{QV,X,loc} \rbr{G, H} = 0$, it may also attain value $+\ns$.

Next, for two generalized processes $Y$ and $Z$ from the space $\mathcal R$ we define
\[
d_{\ns,X,loc} \rbr{Y, Z} := \sum_{N=1}^{\ns} 2^{-N} \overline{\E} \rbr{Y-Z}_{\cdot \wedge \sigma(X,N)}^*.
\]

Now we will deal with relationship between $d_{\ns,X,loc} \rbr{G\cdot X, H\cdot X}$ and $d_{QV,X,loc} \rbr{G, H} $ when $G$ and $H$ are step processes. We have
\[
\int_0^{\sigma(X, N)} \rbr{G_s - H_s}^2 \dd \sbr{X}_s = \sbr{\rbr{G - H} \cdot X_{\cdot \wedge \sigma(X, N)}}
\]
and, by Remark \ref{conservatism}, $\rbr{G - H} \cdot X_{\cdot \wedge \sigma(X, N)}$ is a martingale. Now, applying the BDG inequality we obtain the estimate
\begin{align*}
\overline{\E} \rbr{ \int_0^{\sigma(X, N)} \rbr{G_s - H_s}^2 \dd \sbr{X}_s}^{1/2} & = \overline{\E} \sbr{\rbr{G - H} \cdot X_{\cdot \wedge \sigma(X, N)}}^{1/2} \\ & \ge C_1^{-1}  \overline{\E} \rbr{G\cdot X - H\cdot X}_{\cdot \wedge \sigma(X, N)}^*,
\end{align*}
which yields
\begin{equation} \label{continuity_int}
d_{\ns,X,loc} \rbr{G\cdot X, H\cdot X} \le C_1 \cdot d_{QV,X,loc} \rbr{G, H}.
\end{equation}

For any martingale $X$ the function $d_{\ns,X,loc}$ also satisfies the triangle inequality, thus is a pseudometric (possibly attaining also value $+\ns$) on the space $\mathcal R$. It is easy to see that $d_{\ns,X,loc}\rbr{Y,Z}=0$ does not imply that $Y$ and $Z$ are equal, but for our purposes we should not distinct such processes. Therefore,  we define $Y$ and $Z$ to be equivalent if $d_{\ns,X,loc}\rbr{Y,Z}=0$. This way we obtain the space ${\mathrm G}$ of equivalence classes of generalized processes from the space $\mathcal R$ with respect to the pseudometric $d_{\ns,X,loc}$. For two elements ${\cal Y}$ and ${\cal Z}$ of ${\mathrm G}$ we define their distance (denoted also by $d_{\ns,X,loc}$) by 
\[
d_{\ns,X,loc}\rbr{{\cal Y}, {\cal Z}} := d_{\ns,X,loc}\rbr{Y, Z},
\]
where $Y$ is any element (representative) of class ${\cal Y}$ and $Z$ is any element (representative) of class ${\cal Z}$. The triangle inequality implies that $d_{\ns,X,loc}\rbr{{\cal Y}, {\cal Z}} $ does not depend on the choice of representatives $Y$ and $Z$.
\begin{proposition} \label{lema_conv_d_inf}
$d_{\ns,X,loc}$ is a metric on the space ${\mathrm G}$ (possibly attaining also value $+\ns$) and ${\mathrm G}$ equipped with this metric is complete. If $Y$ is a generalized process, $Y^n \in \mathcal R$, $n\in \N$, and $\sum_{n=1}^{+\ns} d_{\ns,X,loc} \rbr{{ Y}^n, { Y}} < +\ns$ then the generalized processes $Y^n$ converge to $Y$ locally uniformly w.i.e. Moreover, if $Y^n$, $n\in \N$, are processes then the class ${\cal Y}$, which is the limit of the classes ${\cal Y}^n$ containing $Y^n$ resp., contains a process, thus as a representative $Y$ of ${\cal Y}$ we can take a process. \end{proposition}
\begin{proof}
We start with the proof of the second statement of the thesis. Let $B\subseteq[0,+\ns)\times\Omega$ be the set where all the processes $Y^n$, $n \in \N$, are real, but the sequence of generalized processes $Y^n$ does not converge locally uniformly to $Y$, that is $\rbr{Y^n (\omega) - Y(\omega)}_t^* \nrightarrow 0$. Next, for  $(t, \omega) \in [0,+\ns)\times\Omega$ let $N(t, \omega)$ be the smallest element of $\N \cup \cbr{+\ns}$ such that $N(t, \omega) \ge 1 +  \sup_{0\le s \le t} |X_s (\omega)|$. $N(t, \omega)$ is finite w.i.e. (for example it is finite for all pairs $(t, \omega)$ such that the trajectory $[0,t] \ni s \mapsto X_s(\omega)$ is real and continuous) and $\cbr{(t, \omega) \in [0,+\ns)\times \Omega: N(t, \omega) = +\ns}$ is instantly blockable. For $(t, \omega) \in \tilde{B} =  B\cap \cbr{(t, \omega) \in [0,+\ns)\times \Omega: N(t, \omega) < +\ns}$ and $N \ge N(t, \omega)$ one has
\[
\sum_{n=1}^{+\ns} \rbr{Y^n (\omega) - Y(\omega)}_{\sigma(X(\omega), N)}^*  \ge \sum_{n=1}^{+\ns} \rbr{Y^n (\omega)- Y(\omega)}_t^* =  + \ns.
\]
As a result, for any $\varepsilon >0$, we get 
\[
\varepsilon \sum_{n=1}^{+\ns} \sum_{N=1}^{+\ns} 2^{-N} \rbr{Y^n (\omega) - Y(\omega)}_{\sigma(X(\omega), N)}^* = +\ns.
\]
On the other hand, since 
\begin{align*}
& \overline{\E} \sum_{n=1}^{+\ns} \sum_{N=1}^{+\ns} 2^{-N} \rbr{Y^n (\omega) - Y(\omega)}_{\cdot \wedge \sigma(X(\omega), N)}^* \\
& \le  \sum_{n=1}^{+\ns}  \overline{\E} \sum_{N=1}^{+\ns} 2^{-N} \rbr{Y^n (\omega) - Y(\omega)}_{\cdot \wedge \sigma(X(\omega), N)}^* \\
& = \sum_{n=1}^{+\ns} d_{\ns,X,loc} \rbr{Y^n, Y} = : M < +\ns,
\end{align*}
we know that there exist a nonnegative supermartingale which starts from a capital no greater than $\varepsilon M$ and attains value $+\ns$ on $\tilde{B}$. Since $\varepsilon$ is arbitrary positive real, $\tilde{B}$ is  instantly blockable and the same applies to $B$ since $B \subseteq \tilde{B}\cup \cbr{(t, \omega) \in [0,+\ns)\times \Omega: N(t, \omega) = +\ns}$. 

The proof that $d_{\ns,X,loc}$ defines a metric is omitted. To prove the completeness let $\rbr{{\cal Y}^n}$, ${\cal Y}^n \in {\mathrm G}$ for $n \in \N$, be a Cauchy sequence with respect to $d_{\ns,X,loc}$. Let $\rbr{d_k}$ be any sequence of positive reals such that  $\sum_{k=1}^{+\ns} d_k  < +\ns$. There exists a subsequence $\rbr{{\cal Y}^{n_k}}$ such that for  $n \ge n_k$, $n,k =1,2,\ldots$ one has $d_{\ns,X,loc}\rbr{{\cal Y}^{n},{\cal Y}^{n_{k}}} \le d_k$. Let $Y^n \in {\cal Y}^n$, $n \in \N$, be a representative of the class ${\cal Y}^n$ and let $Y := \liminf_{l \ra + \ns} Y^{n_l}$ on the set where all the processes $Y^n$, $n \in \N$, are real. We have  
\begin{align*}
d_{\ns,X,loc}\rbr{Y^{n},Y} & \le d_{\ns,X,loc}\rbr{Y^{n},Y^{n_k}} +  \sum_{l=k}^{+\ns} d_{\ns,X,loc}\rbr{Y^{n_l},Y^{n_{l+1}}} \\
& \le d_k + \sum_{l=k}^{+\ns} d_l.
\end{align*}
Thus, from the already proven second statement of the thesis we have that ${Y}^{n_k}$ converge to $Y$ locally uniformly w.i.e. and, as a result, $Y \in \cal R$ and the classes ${\cal Y}^n$  converge to the class ${\cal Y}$  containing $Y$. 

The fact that if $Y^n$ are processes then as a representative $Y$ of ${\cal Y}$ we can take also a process, follows from the proof of completeness, more precisely from the fact that as the limit $Y$ one may take $\liminf$ of some subsequence of $\rbr{Y^n}$. 
\end{proof}

What will be important to us is that if the filtration $\mathbb{F}$ is right-continuous then for any real process $F$ with c\`{a}dl\`{a}g trajectories, which is globally bounded ($\sup_{(t,\omega) \in [0, +\ns) \times \Omega} |F_t(\omega)| <+\ns$), we are able to construct a sequence of simple trading strategies $\tilde{F}^m=\rbr{0,\rbr{\tau_{n}^m}_n,\rbr{f_{n}^m}_n}$, $m \in \N$, such that the sequence $\rbr{F^m}$ of step processes $$F^m_t := \sum_{n=1}^{+\ns} f^m_{n-1} {\bf 1}_{\left[\tau^m_{n-1}, \tau^m_n \right)} (t)$$ converges in $d_{QV,X,loc}$ to $F$ for all $X$, $\lim_{m \ra +\ns} d_{QV,X,loc}(F^m, F) = 0$. For example, we can define $\tau_0^m := 0$,
\[
\tau_n^m := \inf \cbr{t > \tau_{n-1}^m : |F_t - F_{\tau_{n-1}^m}| \ge  2^{-m}}, \quad n = 1,2,\ldots
\]
and $f_{n}^m = F_{\tau_{n}^m}$. For any $t \in [0, +\ns)$ we naturally have $|F_t - F_t^m| \le  2^{-m}$. The assumption that the process $F$ is c\`{a}dl\`{a}g and the filtration $\mathbb{F}$ is right-continuous guarantees that $\tau_n^m$ are indeed stopping times.

For a simple trading strategy $\tilde{F}$ and its corresponding step process $F$, instead of $\tilde{F} \cdot X$ we will often write ${F} \cdot X$.

Now, using It\^{o}'s isometry we estimate
\begin{align}
d_{QV,X,loc}(F^m, F) & = \sum_{N=1}^{\ns} 2^{-N} \overline{\E} \rbr{ \int_0^{\sigma(N)} \rbr{F^m_s - F_s}^2 \dd \sbr{X}_s}^{1/2} \nonumber \\
& \le \sum_{N=1}^{\ns} 2^{-N} \overline{\E} 2^{-m}  \sbr{X}_{\sigma(N)}^{1/2} \nonumber \\
& \le 2^{-m} \sum_{N=1}^{\ns} 2^{-N} \overline{\E}  \rbr{\frac{1}{2} \sbr{X}_{\sigma(N)}+\frac{1}{2} } \nonumber \\
& \le 2^{-m} \sum_{N=1}^{\ns} 2^{-N}  \rbr{\frac{1}{2} 4 N^2+\frac{1}{2} } = 12.5 \cdot 2^{-m}. \label{ala}
\end{align}

Using \eqref{continuity_int}, \eqref{ala} and the fact that $d_{QV,X,loc}(F^m, F^n) \le d_{QV,X,loc}(F^m, F) + d_{QV,X,loc}(F^n, F) $ we obtain that the sequence of classes whose sequence of representatives is $\rbr{F^m \cdot X}$, is a Cauchy sequence in the space of equivalence classes of generalized processes ${\mathrm G}$, equipped with the metric $d_{\ns,X,loc}$. 

Now, using Proposition \ref{lema_conv_d_inf} we are able to extend the definition of the integral $F \cdot X$ at least for any adapted, globally bounded, real process $F$ with c\`{a}dl\`{a}g trajectories. 
\begin{definition} \label{integrrr}
For any real process $F$, for which there exists a sequence of step processes $\rbr{F^m}$ such that $\lim_{m \ra +\ns } d_{QV,X,loc}(F^m, F) =0$ we define  \emph{the model-free integral} $F \cdot X$ as any process which is a representative of the limit of the (classes containing) integrals $F^m \cdot X$, $m \in \N$,  in the space ${\mathrm G}$ equipped with the metric $d_{\ns,X,loc}$.

If the filtration $\mathbb{F}$ is right-continuous then for any globally bounded, real process $F$ with c\`{a}dl\`{a}g trajectories by \emph{the model-free integral} $F \cdot X$ we will mean any process which is a representative of the limit of the (classes containing) integrals $F^m \cdot X$, $m \in \N$, where $F^m_t := \sum_{n=1}^{+\ns} f^m_{n-1} {\bf 1}_{\left[\tau^m_{n-1}, \tau^m_n \right)} (t)$ and $\tau_0^m := 0$,
\[
\tau_n^m := \inf \cbr{t > \tau_{n-1}^m : |F_t - F_{\tau_{n-1}^m}| \ge  2^{-m}}, \quad n = 1,2,\ldots, 
\]
and $f_{n}^m = F_{\tau_{n}^m}$,
in the above mentioned space.
\end{definition} 

Since $F^m \cdot X$ are martingales then by Proposition \ref{lema_conv_d_inf} we get that the integral $F \cdot X$ is also a martingale (we may always take a subsequence $\rbr{F^{m_k} \cdot X}_k$ of $\rbr{F^m \cdot X}$ such that $\sum_{k=1}^{+\ns}  d_{QV,X,loc}\rbr{F^{m_k}, F} < +\ns$). Therefore, it is in place to calculate its quadratic variation or more generally, the quadratic covariation of two integrals $F \cdot X$ and $G \cdot Y$. 

\begin{fact} \label{q_covar_int}
Let the filtration $\mathbb{F}$ be right-continuous, $G$ and $H$ be globally bounded, real processes with c\`{a}dl\`{a}g trajectories and $X$, $Y$ be martingales. Then the quadratic covariation of the martingales $G \cdot X$ and $H \cdot Y$ equals 
\[
\sbr{G \cdot X, H \cdot Y}_t = \int_0^t \rbr{G_s \cdot H_s}\cdot \dd [X, Y]_s \text{ w.i.e.}
\]
(the integral $\int_0^t \rbr{G_s \cdot H_s}\cdot \dd [X, Y]_s$ is understood as the (pathwise) Lebesque-Stieltjes integral).
\end{fact}
\begin{proof}
Let us consider $m \in \N$ and let $G^m$ and $H^m$ be step processes  such that $|G_t - G_t^{m}| \le  2^{-m}$ and $|H_t - H_t^{m}| \le  2^{-m}$.

Next, by polarization formula of Subsection \ref{q_cov} we know that 
\begin{align*}
\frac{1}{4}\rbr{ G \cdot X + H \cdot Y}^2 & - \frac{1}{4}\rbr{ G \cdot X - H \cdot Y}^2 - \sbr{G \cdot X, H \cdot Y} \\
&= \rbr{ G \cdot X }\rbr{H \cdot Y}  - \sbr{G \cdot X, H \cdot Y}
\end{align*}
is a martingale, and the same applies to the process $\rbr{ G^m \cdot X }\rbr{H^m \cdot Y}  - \sbr{G^m \cdot X, H^m \cdot Y}$. By  Fact \ref{q_var_simp_int} we have $\sbr{G^m \cdot X, H^m \cdot Y}= \int_0^{\cdot} G^m_s \cdot H^m_s \cdot \dd \sbr{X, Y}_s$ w.i.e.

By Proposition \ref{lema_conv_d_inf} we get that $G^m \cdot X$ and $H^m \cdot Y$ tend locally uniformly w.i.e. to $G \cdot X$ and $H \cdot Y$ respectively, as $m \ra +\ns$, and it is easy to see that $\sbr{G^m \cdot X, H^m \cdot Y} = \int_0^{\cdot} G^m_s \cdot H^m_s \cdot \dd \sbr{X, Y}_s$ tends to $\int_0^{\cdot} G_s \cdot H_s  \cdot \dd \sbr{X, Y}_s$ locally uniformly w.i.e. as $m \ra +\ns$. Thus, the differences
\[
 \rbr{ G \cdot X }\rbr{H \cdot Y}  - \sbr{G \cdot X, H \cdot Y} - \rbr{\rbr{ G^m \cdot X }\rbr{H^m \cdot Y} -  \int_0^{\cdot} G^m_s \cdot H^m_s \cdot \dd \sbr{X, Y}_s} 
\]
tend locally uniformly w.i.e. as $m \ra +\ns$ to a martingale which is equal 
\[
Z_t(\omega) = \int_0^t G_s(\omega) \cdot H_s(\omega)  \cdot \dd \sbr{X, Y}_s(\omega) - \sbr{G \cdot X, H \cdot Y}_t(\omega).
\]
This martingale has finite total variation and continuous trajectories on any interval $[0,t]$ w.i.e. and, by standard arguments, its quadratic variation $[Z]$ vanishes w.i.e. Thus, by  Fact \ref{q_var_mart0}, $Z = 0$ w.i.e. which is equivalent with the fact that
\[
\sbr{G \cdot X, H \cdot Y}= \int_0^{\cdot} G_s \cdot H_s  \cdot \dd \sbr{X, Y}_s \text{ w.i.e. }
\]
\end{proof}
 
The assumption in the second part of Definition \ref{integrrr} that the c\`{a}dl\`{a}g process $F$ is globally bounded seems to be too restrictive, therefore now we extend the definition of $F\cdot X$ by localization. For $N>0$ let $\sigma\rbr{F,N}$ be defined similarly as $\sigma(X,N)$ defined by (\ref{eq:sigmaM}). Let us consider the stopped processes $F^N_t(\omega) := F_{t\wedge \sigma\rbr{F,N}}(\omega)$, $N \in \N$. Using Fact \ref{q_covar_int} for any $t\ge 0$ we get 
\begin{align*}
& \sbr{ \rbr{F^{N+1} \cdot X - F^N \cdot X}_{\cdot\wedge \sigma\rbr{F,N}}}_t = \sbr{ {\rbr{F^{N+1} - F^N} \cdot X}}_{t\wedge \sigma\rbr{F,N}}  \\
& = \int_0^{t\wedge \sigma\rbr{F,N}} \rbr{F^{N+1}_s - F^N_s}^2 \dd [X]_s = 0
\end{align*}
since $F^{N+1}_s = F^N_s$ for $s \in \sbr{0,  t\wedge \sigma\rbr{F,N}} \subseteq \sbr{0, \sigma\rbr{F,N}}$. Therefore, by Fact \ref{q_var_mart0},  
both integrals -- $F^{N+1} \cdot X$ and $F^N \cdot X$ -- coincide w.i.e. for $t \in \sbr{0, \sigma\rbr{F,N}}$ and we can define the integral $F \cdot X$ as
\[
F \cdot X = \liminf_{N \ra +\ns} F^{N} \cdot X.
\]

\section{Quadratic variation expressed via limit of truncated variations} 

In this section, for any martingale $X$ we present another sequence of processes which tend locally uniformly w.i.e. to the quadratic variation of $X$. 
To define these processes we introduce \emph{truncated variation} of a c\`{a}dl\`{a}g function $x:[0,+\ns) \ra \R$. The truncated variation of $f$ over the interval $[a,b] \subset [0, +\ns)$ ($-\ns < a < b < +\ns$) with the truncation parameter $c>0$ is defined as
\[
\TTV{x}{\left[a,b\right]}c:=\sup_{n}\sup_{a\le t_{0}<t_{1}<\ldots<t_{n}\le b}\sum_{i=1}^{n}\max\left\{ \left|x\left(t_{i}\right)-x\left(t_{i-1}\right)\right|-c,0\right\} .
\]
Notice that $\TTV{x}{\left[a,b\right]}c$ does not depend on any partition, since it is the supremum over \emph{all} partitions of the interval $[a, b]$.

\begin{proposition}
Let $X$ be a martingale and $\rbr{c_n}$ a sequence of positive reals tending to $0$. The processes $t \mapsto c_n \cdot \TTV{X}{\left[0,t\right]}{c_n}$ tend locally uniformly w.i.e. to $\sbr{X}$ as $n \ra +\ns$. 
\end{proposition}
\begin{proof}[Sketch of a proof]
First we will prove the thesis for the sequence of processes $t \mapsto m^{-2} \cdot \TTV{X}{\left[0,t\right]}{m^{-2}}$, $m=1,2,\ldots$.
Let $M$ be a positive real and $\sigma(M)=\sigma(X,M)$ be
defined by (\ref{eq:sigmaM}). Let $\tau{}^{m,k}=\rbr{\tau{}_{n}^{m,k}}_{n}:=\tau(X,m^{-2},k\cdot m^{-3})$,
$m\in\N\setminus\cbr {0,1}$, $k\in\cbr{0,1,2,\ldots,m-1}$, be the Lebesgue
sequence of stopping times for $X$ and the grid $m^{-2}\cdot\Z+k\cdot m^{-3}$.
We define $\tau^{m,k}\wedge\sigma(M)$ as the sequence $\rbr{\tau_{n}^{m,k}\wedge\sigma(M)}_{n}$.
For $m\in\N\setminus\cbr 0$, $k\in\cbr{0,1,2,\ldots,m-1}$, let $\upsilon^{m,k}$
be the non-decreasing rearrangement of the stopping times from both
sequences $\tau{}^{m,k}$ and $\tau{}^{m,0}$ with redundancies deleted.
$\upsilon^{m,k}=\rbr{\upsilon_{l}^{m,k}}_{l}$ is a proper sequence
of stopping times and we define $\upsilon^{m,k}\wedge\sigma(M)$ as
the sequence $\rbr{\upsilon_{l}^{m,k}\wedge\sigma(M)}_{l}$. Similarly
as in the proof of Fact \ref{103} (inequalities (\ref{eq:ineqlan}) and
(\ref{eq:ineqtwo})), we infer that 
\[
\overline{\E}\rbr{\sbr X^{\tau^{m,k}\wedge\sigma(M)}-\sbr X^{\upsilon^{m,k}\wedge\sigma(M)}}^{*}\le6m^{-2}\rbr{1+4M^{2}}+2m^{-4}
\]
and
\[
\overline{\E}\rbr{\sbr X^{\tau^{m,0}\wedge\sigma(M)}-\sbr X^{\upsilon^{m,k}\wedge\sigma(M)}}^{*}\le6m^{-2}\rbr{1+4M^{2}}+2m^{-4}
\]
thus 
\begin{equation} \label{wie_conv_k}
\overline{\E}\rbr{\sbr X^{\tau^{m,k}\wedge\sigma(M)}-\sbr X^{\tau^{m,0}\wedge\sigma(M)}}^{*}\le12m^{-2}\rbr{1+4M^{2}}+4m^{-4}.
\end{equation}
Summing both sides of \eqref{wie_conv_k} over $k \in\cbr{0,1,2,\ldots,m-1}$ and dividing by $m$ we get that 
\[
\overline{\E} \rbr{\frac{1}{m}\sum_{k=0}^{m-1}\sbr X^{\tau^{m,k}\wedge\sigma(M)}-\sbr X^{\tau^{m,0}\wedge\sigma(M)}}^{*}\le12m^{-2}\rbr{1+4M^{2}}+4m^{-4}
\]
which yields
\begin{equation} \label{wie_conv_kk}
\sum_{m=1}^{+\ns} \overline{\E} \rbr{\frac{1}{m}\sum_{k=0}^{m-1}\sbr X^{\tau^{m,k}\wedge\sigma(M)}-\sbr X^{\tau^{m,0}\wedge\sigma(M)}}^{*} < +\ns.
\end{equation}
Notice that on the set where 
\[
\rbr{\frac{1}{m}\sum_{k=0}^{m-1}\sbr X^{\tau^{m,k}\wedge\sigma(M)}-\sbr X^{\tau^{m,0}\wedge\sigma(M)}}^{*}\nrightarrow_{m\ra+\ns}0
\]
one has
\[
\sum_{m=1}^{+\ns}\rbr{\frac{1}{m}\sum_{k=0}^{m-1}\sbr X^{\tau^{m,k}\wedge\sigma(M)}-\sbr X^{\tau^{m,0}\wedge\sigma(M)}}^{*}=+\ns.
\]
This and \eqref{wie_conv_kk} imply that $\frac{1}{m}\sum_{k=0}^{m-1}\sbr X^{\tau^{m,k}}$ tends locally uniformly and w.i.e. to the same limit as $\sbr X^{\tau^{m,0}}$, that is to $\sbr X$.

The next ingredient of the proof which we need is the following identity
\begin{equation} \label{banach_ind}
\TTV X{[0,\sigma(M)\wedge t]}{m^{-2}}=\int_{\R}n^{z,m^{-2}}(X,[0,\sigma(M)\wedge t])\dd z,
\end{equation}
where for a c\`{a}dl\`{a}g function $x:[0,+\ns) \ra \R$ and real numbers $0 \le a<b <+\ns$, $c>0$, $n^{z,c}(x,[a,b])$ denotes the number of crossings by $x$ the value interval $[z -c/2, z+c/2]$ on the interval $[a,b]$. For precise definitions of $n^{z,c}(x,[a,b])$ see \cite[Subsect. 2.4]{LochOblPS:2021} and for the proof of \eqref{banach_ind} see \cite{LochowskiColloquium:2017}. 

Next, let us notice that for $t>0$, $m\in\N\setminus\cbr {0,1}$ and $k\in\cbr{0,1,2,\ldots,m-1}$
\begin{equation} \label{q_var_leb}
\sbr X^{\tau^{m,k}\wedge\sigma(M)}_t = \sum_{p \in \Z} m^{-4} n^{m^{-2} \cdot p + m^{-2}/2 + k\cdot m^{-3} , m^{-2}}(X,[0,t \wedge \sigma(M)])
\end{equation}
since $X$ has continuous trajectories and the Lebesgue stopping times are hitting times of consecutive levels of the grid $m^{-2} \cdot \Z + k\cdot m^{-3}$.

For $p \in \Z$, $m\in\N\setminus\cbr {0,1,2 }$, $k\in\cbr{0,1,2,\ldots,m-2}$ and 
\begin{equation} \label{z_interval}
z \in \left[ \frac{p}{(m-1)^{2}}+\frac{k}{(m-1)^{3}}, \frac{p}{(m-1)^{2}}+\frac{k+1}{(m-1)^{3}} \right) 
\end{equation}
we have
\begin{align*}
\frac{p}{(m-1)^{2}}+\frac{k}{(m-1)^{3}}+\frac{1}{m^{2}} & \le z+\frac{1}{m^{2}}<\frac{p}{(m-1)^{2}}+\frac{k+1}{(m-1)^{3}}+\frac{1}{m^{2}}\\
 & <\frac{p+1}{(m-1)^{2}}+\frac{k}{(m-1)^{3}},
\end{align*}
which follows from the estimate
\begin{align*}
& \frac{p+1}{(m-1)^{2}}+\frac{k}{(m-1)^{3}}-\frac{p}{(m-1)^{2}}-\frac{k+1}{(m-1)^{3}}-\frac{1}{m^{2}} \\ & =\frac{1}{(m-1)^{2}}-\frac{1}{(m-1)^{3}}-\frac{1}{m^{2}}
 =\frac{m^{2}-3m+1}{m^{2}(m-1)^{3}}>0,
\end{align*}
valid for $m\ge3$. Thus, each crossing of the interval 
\[
\sbr{ \frac{p}{(m-1)^{2}}+\frac{k}{(m-1)^{3}}, \frac{p+1}{(m-1)^{2}}+\frac{k}{(m-1)^{3}} }
\]
implies crossing of the interval $\sbr{z, z+m^{-2}}$, whenever $z$ satisfies \eqref{z_interval}.
This implies that 
\begin{align*}
& \int_{ \frac{p}{(m-1)^{2}}+\frac{k}{(m-1)^{3}}}^{\frac{p}{(m-1)^{2}}+\frac{k+1}{(m-1)^{3}}} n^{z + m^{-2}/2, m^{-2}}(X,[0,t \wedge \sigma(M)]) \dd z \\& \le\int_{ \frac{p}{(m-1)^{2}}+\frac{k}{(m-1)^{3}}}^{\frac{p}{(m-1)^{2}}+\frac{k+1}{(m-1)^{3}}}  n^{p{(m-1)^{-2}}+{k}{(m-1)^{-3}} + (m-1)^{-2}/2, (m-1)^{-2} } (X,[0,t \wedge \sigma(M)]) \dd z \\
& = \frac{1}{(m-1)^3} n^{p{(m-1)^{-2}}+{k}{(m-1)^{-3}} + (m-1)^{-2}/2, (m-1)^{-2} } (X,[0,t \wedge \sigma(M)]).
\end{align*}
Now, summing over $p \in \Z$ and $k\in\cbr{0,1,2,\ldots,m-2}$, and using \eqref{q_var_leb} (with $m$ replaced by $m-1$) we get 
\begin{align}
& \TTV X{[0,\sigma(M)\wedge t]}{m^{-2}} \nonumber \\
& = \int_{\R} n^{z + m^{-2}/2, m^{-2}}(X,[0,t \wedge \sigma(M)]) \dd z \nonumber \\
& = \sum_{p \in \Z} \sum_{k=0}^{m-2} \int_{ \frac{p}{(m-1)^{2}}+\frac{k}{(m-1)^{3}}}^{\frac{p}{(m-1)^{2}}+\frac{k+1}{(m-1)^{3}}} n^{z + m^{-2}/2, m^{-2}}(X,[0,t \wedge \sigma(M)]) \dd z \nonumber \\
& \le \frac{1}{(m-1)^3}  \sum_{k=0}^{m-2} \sum_{p \in \Z}  n^{p{(m-1)^{-2}}+{k}{(m-1)^{-3}} + (m-1)^{-2}/2, (m-1)^{-2} }(X,[0,t \wedge \sigma(M)]) \nonumber \\
& = (m-1) \sum_{k=0}^{m-2} \sbr X^{\tau^{m-1,k}\wedge\sigma(M)}_t. \label{tvqvar_1}
\end{align}

On the other hand, for $p \in \Z$, $m\in\N\setminus\cbr {0}$, $k\in\cbr{0,1,2,\ldots,m}$ and 
\begin{equation} \label{z_interval1}
z \in \left( \frac{p}{(m+1)^{2}}+\frac{k}{(m+1)^{3}}, \frac{p}{(m+1)^{2}}+\frac{k+1}{(m+1)^{3}} \right] 
\end{equation}
we have
\begin{align*}
\frac{p+1}{(m+1)^{2}}+\frac{k+1}{(m+1)^{3}} & <\frac{p}{(m+1)^{2}}+\frac{k}{(m+1)^{3}}+\frac{1}{m^{2}}\\
 & <z+\frac{1}{m^{2}}<\frac{p}{(m+1)^{2}}+\frac{k+1}{(m+1)^{3}}+\frac{1}{m^{2}}
\end{align*}
since
\begin{align*}
& \frac{p}{(m+1)^{2}}+\frac{k}{(m+1)^{3}}+\frac{1}{m^{2}}-\frac{p+1}{(m+1)^{2}}-\frac{k+1}{(m+1)^{3}} \\
& =\frac{1}{m^{2}}-\frac{1}{(m+1)^{2}}-\frac{1}{(m+1)^{3}}  =\frac{m^{2}+3m+1}{m^{2}(m+1)^{3}}>0.
\end{align*}
Thus, each crossing of the interval $\sbr{z, z+m^{-2}}$
implies a crossing of the interval 
\[
\sbr{ \frac{p}{(m-1)^{2}}+\frac{k+1}{(m-1)^{3}}, \frac{p+1}{(m-1)^{2}}+\frac{k+1}{(m-1)^{3}} }
\]
whenever $z$ satisfies \eqref{z_interval1}.
This implies analogous inequality to \eqref{tvqvar_1}, but in opposite direction:
\begin{align}
& \TTV X{[0,\sigma(M)\wedge t]}{m^{-2}} \ge {(m+1)}  \sum_{k=0}^{m}  \sbr X^{\tau^{m+1,k}\wedge\sigma(M)}_t. \label{tvqvar_2}
\end{align}
\eqref{tvqvar_1} and \eqref{tvqvar_2} give bounds
\begin{align*}
\frac{m+1}{m^2}  \sum_{k=0}^{m}  \sbr X^{\tau^{m+1,k}\wedge\sigma(M)}_t & \le \frac{1}{m^2} \TTV X{[0,\sigma(M)\wedge t]}{m^{-2}} \\
& \le \frac{m-1}{m^2}  \sum_{k=0}^{m}  \sbr X^{\tau^{m-1,k}\wedge\sigma(M)}_t,
\end{align*}
which imply that $m^{-2} \TTV X{[0,\cdot]}{m^{-2}}$ tends locally uniformly and w.i.e. to the same limit as $\frac{1}{m}  \sum_{k=0}^{m}  \sbr X^{\tau^{m,k}}$, that is to $\sbr X$.

Finally,  the convergence of $c_{n}\cdot\TTV X{\left[0,\cdot \right]}{c_{n}}$
for any sequence $c_{n}\ra0+$ follows from the estimates 
\begin{align*}
& \frac{\left\lfloor 1/\sqrt{c_{n}}\right\rfloor^2 }{\left\lfloor 1/c_{n}\right\rfloor +1}\frac{1}{\left\lfloor 1/\sqrt{c_{n}}\right\rfloor^2 }\cdot\TTV X{\left[0,t\right]}{1/\left\lfloor 1/\sqrt{c_{n}}\right\rfloor ^2} \\ & \leq c_{n}\cdot\TTV X{\left[0,t\right]}{c_{n}}
 \leq\frac{\left\lceil 1/\sqrt{c_{n}}\right\rceil ^2}{\left\lceil 1/c_{n}\right\rceil -1}\frac{1}{\left\lceil 1/\sqrt{c_{n}}\right\rceil^2 }\cdot\TTV X{\left[0,t\right]}{1/\left\lceil 1/\sqrt{c_{n}}\right\rceil^2}
\end{align*}
valid for $c_{n}<1,$ which stem directly from inequalities 
\[
\frac{1}{\left\lfloor 1/c_{n}\right\rfloor +1}\le c_{n}\le\frac{1}{\left\lceil 1/c_{n}\right\rceil -1}\mbox{ and }\frac{1}{\left\lceil 1/\sqrt{c_{n}}\right\rceil^2 }\le c_{n}\le\frac{1}{\left\lfloor 1/\sqrt{c_{n}}\right\rfloor^2 }
\]
(valid for $c_{n}<1$), and the fact that the function $\left(0,+\ns\right)\ni c\mapsto\TTV X{\left[0,t\right]}c$
is non-increasing.
\end{proof}






{\bf Acknowledgments}
The author would like to thank Vladimir Vovk for his comments on the notion of the outer measure presented in this article. Valuable remarks of an anonymous referee helped to improve the presentation and correct many gaps in the previous version of the article. The author is thankful for this. 
The research presented in this article was partially funded by the research project no. KAE/S21/1.43 \emph{Methods of mathematical economics and their applications in contemporary transformations} of Department of Mathematics and Mathematical Economics, Warsaw School of Economics, and partially by the grants no. 2019/35/B/ST1/04292 \emph{Generic chaining approach to the regularity of stochastic processes} and no. 2022/47/B/ST1/02114 \emph{Non-random equivalent characterizations of sample boundedness} of  National Science Centre, Poland.

\bibliographystyle{amsalpha}
\bibliography{/Users/rafallochowski/biblio/biblio}

\end{document}